\documentclass[twocolumn]{aastex63}

\usepackage[T1]{fontenc} % for Icelandic Characters
\usepackage[utf8]{inputenc} % for Icelandic Characters 

\newcommand\tess{TESS}

\newcommand\ms{$\textrm{m~s}^{-1}$}
\newcommand\cms{$\textrm{cm~s}^{-1}$}

\newcommand\teff{T$_{\rm{eff}}$}

\received{March 25, 2023}
\revised{June 21, 2023}
\accepted{\today}
\newcommand{\NOIRLab}{NSF's National Optical-Infrared Astronomy Research Laboratory, 950 N.\ Cherry Ave., Tucson, AZ 85719, USA}
\newcommand{\GoddardESAL}{Exoplanets and Stellar Astrophysics Laboratory, NASA Goddard Space Flight Center, Greenbelt, MD 20771, USA}

\shortauthors{Kanodia et al. 2023}
\shorttitle{NEID Fiber Feed}

\submitjournal{AJ}

\begin{document}

\title{Stable fiber-illumination for extremely precise radial velocities with NEID}

\author[0000-0001-8401-4300]{Shubham Kanodia}
\affiliation{Department of Astronomy \& Astrophysics, 525 Davey Laboratory, The Pennsylvania State University, University Park, PA, 16802, USA}
\affiliation{Center for Exoplanets and Habitable Worlds, 525 Davey Laboratory, The Pennsylvania State University, University Park, PA, 16802, USA}
\affiliation{Earth and Planets Laboratory, Carnegie Institution for Science, 5241 Broad Branch Road, NW, Washington, DC 20015, USA}

\author[0000-0002-9082-6337]{Andrea S.J.\ Lin}
\affil{Department of Astronomy \& Astrophysics, 525 Davey Laboratory, The Pennsylvania State University, University Park, PA, 16802, USA}
\affil{Center for Exoplanets and Habitable Worlds, 525 Davey Laboratory, The Pennsylvania State University, University Park, PA, 16802, USA}

\author[0000-0003-0790-7492]{Emily Lubar}
\affil{Aerospace Corporation, 2310 E. El Segundo Blvd. El Segundo, CA 90245-4609}

\author[0000-0003-1312-9391]{Samuel Halverson}
\affil{Jet Propulsion Laboratory, California Institute of Technology, 4800 Oak Grove Drive, Pasadena, California 91109}

\author[0000-0001-9596-7983]{Suvrath Mahadevan}
\altaffiliation{NEID Principal Investigator}
\affil{Department of Astronomy \& Astrophysics, 525 Davey Laboratory, The Pennsylvania State University, University Park, PA, 16802, USA}
\affil{Center for Exoplanets and Habitable Worlds, 525 Davey Laboratory, The Pennsylvania State University, University Park, PA, 16802, USA}
\affil{ETH Zurich, Institute for Particle Physics \& Astrophysics, Switzerland}

\author[0000-0003-4384-7220]{Chad F. Bender}
\affil{Steward Observatory, The University of Arizona, 933 N.\ Cherry Avenue, Tucson, AZ 85721, USA}

\author[0000-0002-9632-9382]{Sarah E.\ Logsdon}
\affil{\NOIRLab}

\author[0000-0002-4289-7958]{Lawrence W. Ramsey}
\affil{Department of Astronomy \& Astrophysics, 525 Davey Laboratory, The Pennsylvania State University, University Park, PA, 16802, USA}
\affil{Center for Exoplanets and Habitable Worlds, 525 Davey Laboratory, The Pennsylvania State University, University Park, PA, 16802, USA}

\author[0000-0001-8720-5612]{Joe P.\ Ninan}
\affil{Department of Astronomy and Astrophysics, Tata Institute of Fundamental Research, Homi Bhabha Road, Colaba, Mumbai 400005, India}

\author[0000-0001-7409-5688]{Guðmundur Stefánsson} 
\altaffiliation{NASA Sagan Fellow}
\affil{Department of Astrophysical Sciences, Princeton University, 4 Ivy Lane, Princeton, NJ 08540, USA} 

\author[0000-0002-0048-2586]{Andrew Monson}
\affil{Steward Observatory, The University of Arizona, 933 N.\ Cherry Avenue, Tucson, AZ 85721, USA}

\author[0000-0002-4046-987X]{Christian Schwab}
\affil{School of Mathematical and Physical Sciences, Macquarie University, Balaclava Road, North Ryde, NSW 2109, Australia}

\author[0000-0001-8127-5775]{Arpita Roy}
\affil{Space Telescope Science Institute, 3700 San Martin Dr, Baltimore, MD 21218, USA}
\affil{Department of Physics and Astronomy, Johns Hopkins University, 3400 N Charles Street, Baltimore, MD 21218, USA}

\author[0000-0003-1324-0495]{Leonardo A. Paredes}
\affil{Steward Observatory, The University of Arizona, 933 N.\ Cherry Avenue, Tucson, AZ 85721, USA}

\author{Eli Golub}
\affil{\NOIRLab}

\author[0000-0002-3985-8528]{Jesus Higuera}
\affil{\NOIRLab}

\author[0000-0003-3906-9518]{Jessica Klusmeyer}
\affil{\NOIRLab}

\author{William McBride}
\affil{\NOIRLab}

\author[0000-0002-6096-1749]{Cullen Blake}
\affil{ Department of Physics and Astronomy, University of Pennsylvania,  209 South 33rd Street, Philadelphia, PA 19104, USA}

\author[0000-0002-2144-0764]{Scott A. Diddams}
\affil{Electrical, Computer \& Energy Engineering, University of Colorado, 425 UCB, Boulder, CO 80309, USA}
\affil{Department of Physics, University of Colorado, 2000 Colorado Avenue, Boulder, CO 80309, USA}
\affil{National Institute of Standards and Technology, 325 Broadway, Boulder, CO 80305, USA}

\author[0000-0002-3603-9789]{Fabien Grisé}
\affil{Department of Astronomy \& Astrophysics, 525 Davey Laboratory, The Pennsylvania State University, University Park, PA, 16802, USA}

 \author[0000-0002-5463-9980]{Arvind F.\ Gupta}
\affil{Department of Astronomy \& Astrophysics, 525 Davey Laboratory, The Pennsylvania State University, University Park, PA, 16802, USA}
\affil{Center for Exoplanets and Habitable Worlds, 525 Davey Laboratory, The Pennsylvania State University, University Park, PA, 16802, USA}

\author[0000-0002-1664-3102]{Fred Hearty}
\affil{Department of Astronomy \& Astrophysics, 525 Davey Laboratory, The Pennsylvania State University, University Park, PA, 16802, USA}
\affil{Center for Exoplanets and Habitable Worlds, 525 Davey Laboratory, The Pennsylvania State University, University Park, PA, 16802, USA}

\author[0000-0003-0241-8956]{Michael W.\ McElwain}
\affil{\GoddardESAL}

\author[0000-0002-2488-7123]{Jayadev Rajagopal}
\affil{\NOIRLab}

\author[0000-0003-0149-9678]{Paul Robertson}
\altaffiliation{NEID Instrument Team Project Scientist}
\affil{Department of Physics \& Astronomy, University of California Irvine, Irvine, CA 92697, USA}

\author[0000-0002-4788-8858]{Ryan C. Terrien}
\affil{Carleton College, One North College St., Northfield, MN 55057, USA}

% SK, Andrea, Emily, Sam, Suvrath, Chad, Larry, Joe, Arpita

% Emily Lubar, Andrea Lin, Sam Halverson
% Colin Nitroy, Helen Baran, Marissa Maney
% Larry, Sarah, Emily Hunting, Bill McBride, Dan Li, Jessica Klusmeyer (installation of the telescope fiber)
% Fabien Grise - Grating

% \fi

\correspondingauthor{Shubham Kanodia}
\email{skanodia@carnegiescience.edu}

\begin{abstract}
NEID is a high-resolution red-optical precision radial velocity (RV) spectrograph recently commissioned at the WIYN 3.5 m telescope at Kitt Peak National Observatory, Arizona, USA. NEID has an extremely stable environmental control system, and spans a wavelength range of 380 to 930 nm with two observing modes: a High Resolution (HR) mode at R $\sim$ 112,000 for maximum RV precision, and a High Efficiency (HE) mode at R $\sim$ 72,000 for faint targets.  In this manuscript we present a detailed description of the components of NEID's optical fiber feed, which include the instrument, exposure meter, calibration system, and telescope fibers. Many parts of the optical fiber feed can lead to uncalibratable RV errors, which cannot be corrected for using a stable wavelength reference source. We show how these errors directly cascade down to performance requirements on the fiber feed and the scrambling system. We detail the design, assembly, and testing of each component. Designed and built from the bottom-up with a single-visit instrument precision requirement of 27 \cms{}, close attention was paid to the error contribution from each NEID subsystem. Finally, we include the lab and on-sky tests performed during instrument commissioning to test the illumination stability, and discuss the path to achieving the instrumental stability required to search for a true Earth twin around a Solar-type star. 
\end{abstract}


\section{Introduction} \label{sec:intro}

Introduced in astronomical instrumentation in the late 1970s to replace slit illumination, optical fibers were soon recognized for their utility\footnote{See \cite{minardi_astrophotonics_2021} for a review of photonics in astronomical instrumentation.} and scrambling properties \citep[desensitizing the output illumination from changes in the input due to seeing and pointing variations;][]{angel_very_1977, hubbard_operation_1979, serkowski_fabry-perot_1979, black_assessment_1980, barden_evaluation_1981, heacox_application_1986}. Their multiplexing ability made them attractive both for observing multiple objects simultaneously \citep{hill_multiple_1980, hill_history_1988}, as well as for observing different regions of a spatially resolved object \citep{vanderriest_fiber-optics_1980}. One of the first fiber-fed spectrographs was developed at Pennsylvania State University \citep{ramsey_telescope_1980, ramsey_penn_1985}---initially for the 1.5 m at Black Moshannon Observatory, and then the 2.1 m at Kitt Peak National Observatory---for spectral monitoring of RS CVn systems \citep{huenemoerder_fiber-optic-echelle-ccd_1989, hall_fiber-optic_1990}.

Optical fibers were first used for radial velocity (RV) searches for extra-solar planets as part of the ELODIE instrument \citep{baranne_elodie_1996}, which was the successor to CORAVEL \citep{baranne_coravel_1979}. Using the cross correlation technique and a two-fiber input (for simultaneous science and calibration with a thorium argon lamp), ELODIE achieved precision of $\sim$ 13 m/s, which enabled the discovery of the first exoplanet orbiting a solar type star \citep{mayor_jupiter-mass_1995}. In addition to their scrambling abilities, optical fibers enable the RV spectrograph to be decoupled from the telescope prime focus, enabling a static gravitational vector (one that does not change with telescope position), as well as bigger and bulkier instruments with extensive environmental stabilization \citep{pepe_harps_2002, stefansson_versatile_2016, robertson_ultrastable_2019}. The High Accuracy Radial velocity Planet Search (HARPS) spectrometer at La Silla in Chile \citep[][]{mayor_setting_2003} was one of the first spectrographs with environmental stability control, and had the optical bench enclosed inside a vacuum chamber. It also used optical fibers for simultaneous wavelength calibration spanning a large part of the spectrum, enabling it to reach an instrumental precision of 1--3 \ms{} \citep{lovis_exoplanet_2006}.

\begin{figure*}[!ht] 
\center
\includegraphics[width=0.75\textwidth]
{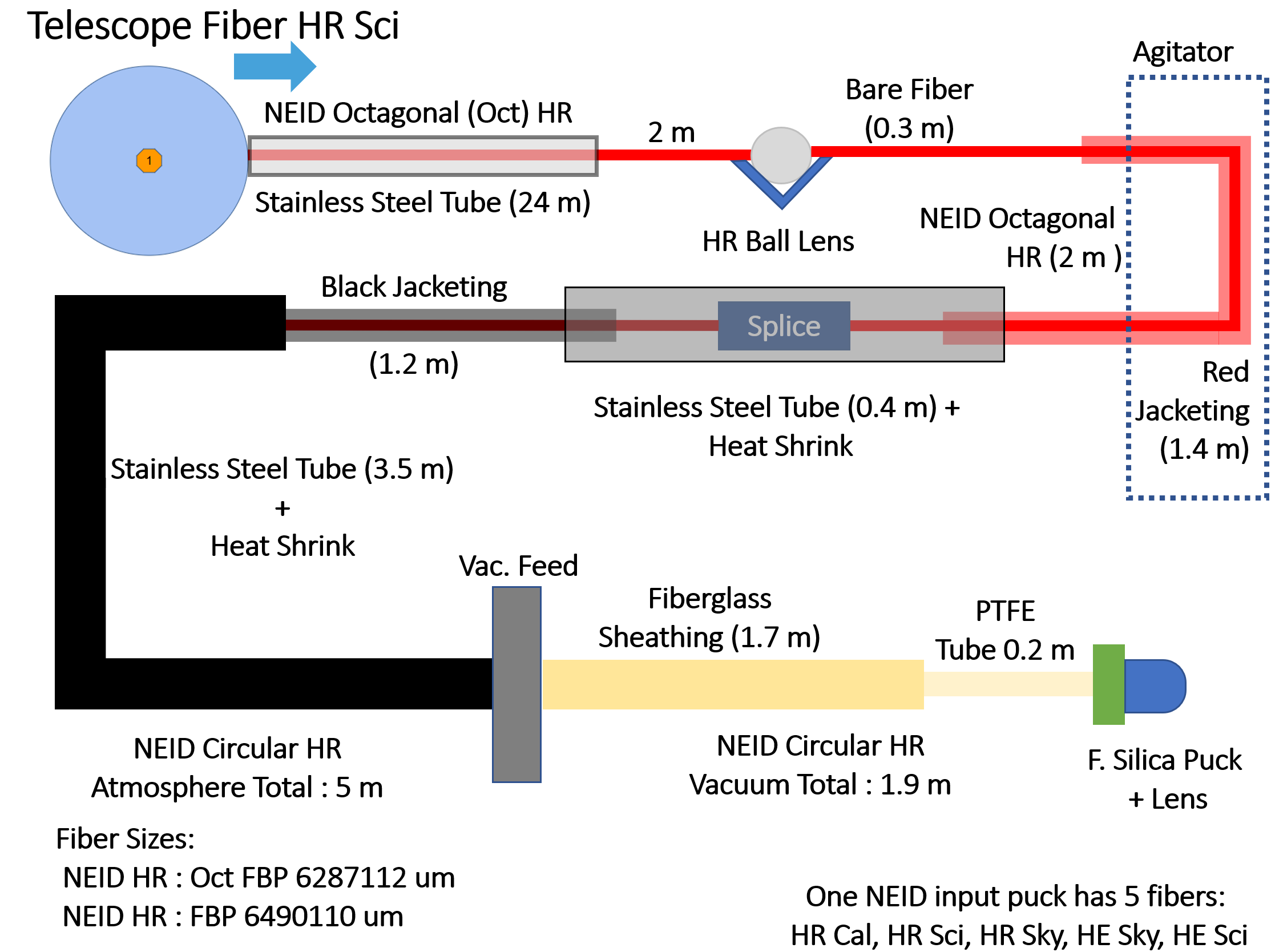}
\caption{Schematic of the NEID fiber train for the HR science fiber (not to scale).} \label{fig:schematic}
\end{figure*}

Precision RV instruments are essential for measuring the masses of planets, especially the low-mass terrestrial ones. For example, the Doppler semi-amplitude of an Earth mass planet orbiting at 1 AU around a Solar mass star is only $\sim 9$ \cms{}, the detection of which requires RV measurement precision at the few \cms{} level. When coupled with transiting surveys such as \textit{Kepler} \citep{borucki_keplers_2009} and \tess{} \citep{ricker_transiting_2014}, RV-derived masses are important for density estimation of exoplanets, which in turn puts bounds on planetary composition. Precise masses are also important to place informative priors on atmospheric scale heights, which are used to estimate the S/N of transmission spectroscopy observations to study planetary atmospheres. RV measurements are also required to obtain orbital parameters for transiting and directly imaged planets, which are crucial for upcoming direct imaging missions such as the Habitable Worlds Observatory \citep{the_luvoir_team_luvoir_2019, gaudi_habitable_2019}. These science objectives necessitate instruments with RV precision at the sub-\ms{} level pushing towards \cms{}.

In this paper we discuss the steps followed in the design, testing and commissioning of the optical fiber train for NEID\footnote{NEID comes from the O’odham word ‘neid’ meaning “to see/visualize.” Kitt Peak National Observatory is located on the Tohono O’odham Nation.}, a new ultra-precise RV spectrograph, installed at the WIYN 3.5 m telescope\footnote{The WIYN Observatory is a joint facility of the NSF’s National Optical-Infrared Astronomy Research Laboratory,
Indiana University, the University of Wisconsin-Madison, Pennsylvania State University, the University of Missouri, the
University of California-Irvine, and Purdue University.} at the  Kitt Peak National Observatory, USA. NEID builds upon the successful extreme environmental stabilization of the near infrared (NIR) Habitable-zone Planet Finder \citep[HPF;][]{mahadevan_habitable-zone_2014, stefansson_versatile_2016, robertson_ultrastable_2019}, and includes a high throughput optical design \citep{schwab_design_2016} and a laser frequency comb (LFC) and a Fabry-P{\'e}rot etalon as calibrators. Designed to achieve a single-visit RV precision of 27 \cms{} \citep{halverson_comprehensive_2016}, NEID is part of the new generation of extreme precision optical RV instruments which include ESPRESSO \citep{pepe_espressovlt_2020}, EXPRES \citep{blackman_performance_2020}, KPF \citep{gibson_kpf_2016} and MAROON-X \citep{seifahrt_development_2016}, among others. 

All these new RV instruments are fiber-fed and utilize a stable calibration source through the Science (or separate Calibration fiber), such as an LFC or etalon cavity, which helps calibrate out instrumental errors due to thermo-mechanical (temperature, pressure, flexing) or detector effects \citep{halverson_comprehensive_2016}. However, crucially, the error contributions from fiber and illumination sources cannot be calibrated out using calibration exposures, and it is therefore imperative to understand and mitigate these error sources. In this manuscript, we detail the various sub-systems of the NEID fiber train, as well as the steps followed for their fabrication. NEID's optical fiber train is conceptually based on that of HPF \citep{kanodia_overview_2018}, but has been modified for the WIYN telescope (diameter and seeing), NEID's higher spectral resolution, and its tighter precision requirements. This detailed description of the NEID fiber-feed is meant to serve as an example for future instruments, but also to highlight the current state of the field, and the pathway (including roadblocks) towards the next generation of RV instrumentation.

The paper is structured as follows: in Section \ref{sec:overview} we give an overview of the NEID fiber feed, while we discuss the design choices for various sub-components in Section \ref{sec:designchoices}. In Section \ref{sec:telescope} we detail the telescope fiber, and the testing carried out to characterize it. In Sections \ref{sec:spatialscrambling} and \ref{sec:instrument} we discuss the various instrument fiber sub-components, while in Section \ref{sec:testing} we describe our testing protocol for each stage in the fiber train. We then discuss our on-sky tests to characterize the scrambling performance of the fiber train during commissioning in Section \ref{sec:onsky}. Finally, in Section \ref{sec:roadmap} we summarize these results and discuss future prospects.

% High level of control on system parameters. 
% Connect to error budget
% Fibers are responsible for stellar light and cal light
% Seeing variations - feed into guiding as well 
% Modal noise

\section{Overview of NEID Fiber Train}\label{sec:overview}
NEID uses a port adapter to interface between the WIYN telescope focus and the input fiber \citep{logsdon_neid_2018, logsdon_neid_2022}.  The port adapter converts the native f/6.3 focal ratio of the WIYN telescope, to the f/4 input of the instrument \citep{schwab_neid_2018}, and it corrects for atmospheric dispersion over the entire wavelength bandpass of NEID (380-930 nm), while providing stable light injection to the fibers using a tip-tilt stage \citep{2018SPIE10702E..6MP, li_neid_2022}. The port adapter was intentionally designed to allow light from the calibration system and solar feed \citep{lin_observing_2022} to follow the same optical path as the stellar data and be projected onto the Science fibers. This system is used to input light from the solar telescope to the spectrograph for RV monitoring of the Sun in the daytime, in order to obtain an independent handle on instrument systematics and also test new stellar activity mitigation techniques on a well-studied stellar system.

\begin{table*}[ht]
\caption{Summary of the dimensions of fiber types used in NEID \label{tab:fibersizes}}
\begin{center}       
\begin{tabular}{cccc}
\hline \hline
Name & Fiber Shape & Fiber dimensions & Application \\
     &          & Core, Cladding, Buffer ($\mu$m) & \\
\hline
Telescope HR & Octagonal & 61.4, 88.5, 112 & Port adapter to Ball scrambler \\
Telescope HE & Octagonal & 101.1, 141.3, 166 & Port adapter to Ball scrambler \\
Instrument HR & Circular & 64.1, 90.2, 108.4 & Ball scrambler to Instrument \\
Instrument HE & Circular & 101.2, 141.2, 168.9 & Ball scrambler to Instrument \\
Exposure Meter & Circular & 150, 165, 195 & Pick off mirror focus to Exposure Meter \\
Solar & Circular & 102, 122, 145 & Solar telescope to Calibration bench \\
Bifurcated Fiber & Octagonal & 101, 141, 166 & Calibration bench$^{a}$ to Port adapter \\
\hline 
\end{tabular}
\tablenotetext{a}{The solar telescope feeds into a shutter tube in the NEID calibration system \citep[Figure 9 from][]{lin_observing_2022}. The bifurcated fiber includes inputs from the solar shutter tube and NEID calibration system integrating sphere that can then be controlled through separate bi-stable shutters.}
\end{center}
\end{table*} 

NEID has two observing modes utilizing two separate fiber feeds, the High Resolution (HR) mode with a spectral resolution of  R $\sim$ 112,000 for precise RV observations, and a High Efficiency (HE) mode for fainter objects, with a spectral resolution of R $\sim$ 72,000. The physical dimensions of the various fibers used in NEID are given in \autoref{tab:fibersizes}. The on-sky sizes of these fibers are 0.92\arcsec{} for HR mode and about 1.5\arcsec{} for HE mode. NEID has three instrument fibers for HR mode: Science, Sky and Calibration. The star light is fed into the Science fiber, while the Sky fiber is used to measure sky brightness and emission lines, and the Calibration fiber is used for simultaneous calibration during science exposures. The HE mode has a Science and Sky fiber, and does not include a simultaneous Calibration fiber. 

The port adapter directs light into either the HR or HE telescope fiberhead by inserting/removing a fold mirror. The fiberhead consists of a central Science octagonal fiber, surrounded by three Sky octagonal fibers separated by 120$^{\circ}$, as well as three Coherent Fiber Bundles (CFBs). The telescope fiber is then fed down from the port adapter to the instrument room. Here the octagonal Science and Sky fibers are fed into a v-groove block where they are face-coupled to the ball lens double scrambler \citep{halverson_efficient_2015}. The output of the ball lens is face-coupled to a 2 meter octagonal fiber, which is then fed to a modal noise agitator for temporal scrambling of the transverse modes propagating through the fiber \citep{roy_scrambling_2014, mahadevan_suppression_2014}. The 2 meter octagonal fiber is spliced on to the circular instrument fiber, which serves as the spectrograph input. The complete instrument fiber bundle consists of the 3 HR (Science, Sky and Calibration), and 2 HE (Science and Sky) fibers, which enter the instrument vacuum chamber through a vacuum feedthrough and are then terminated in a fused silica fiber puck (\autoref{fig:schematic}).  

NEID also has an extensive instrument calibration system, which will be described in detail in a future manuscript, but we include a brief description here. It consists of stable wavelength sources, namely an LFC and an etalon cavity illuminated by a white-light super-continuum source. There is also a Laser-Driven Light Source (LDLS; Hamamatsu EQ-99X) for obtaining \edit1{spectral} flats. These light sources (along with a thorium argon \edit1{oxide} lamp for wavelength calibration), are mounted on a turret and can be selected using an elliptical mirror mounted on a rotation stage \citep[similar to that for HPF;][]{halverson_habitable-zone_2014}. The light is then fed into a 1-inch integrating sphere with two output arms that have separate shutters. One of these outputs feeds directly into the instrument Calibration fiber, whereas the other feeds the bifurcated fiber to the port adapter, allowing it to illuminate the Science fiber and thus follow the same path as the stellar photons (albeit not contemporaneously). As previously mentioned, the solar telescope can also illuminate the Science fiber via the bifurcated fiber, allowing for simultaneous Calibration light with solar observations \citep[Figure 9 in ][]{lin_observing_2022}. 

Lastly, NEID also uses a chromatic exposure meter which picks off a small fraction of the light reflected from the blank between the two mosaics of the echelle grating. This is then re-directed outside the instrument to a separate low-resolution $R \sim$ 100 spectrograph from Wasatch Photonics for faster cadence feedback and monitoring of the flux on the NEID detector \citep{gupta_real-time_2022}. This system allows for chromatic flux-weighted midpoints \citep{blackman_measured_2019} which are required to meet the barycentric correction budget for NEID.

\section{Design Choices}\label{sec:designchoices}
The single-visit instrumental precision for any RV observation is a combination of photon noise, instrumental noise, and astrophysical noise due to Keplerian signals and stellar activity. Astrophysical noise sources are specific to individual stars and beyond the scope of this manuscript. In the subsequent sections we discuss the various design choices motivated by minimizing the photon and instrumental noise contributions for NEID visits.

\subsection{Contributing to Photon Noise}
\subsubsection{Fiber Size} \label{sec:fiber_size}
The RV photon noise (or quality factor) benefits from high spectral resolution, up until it plateaus at about 100,000--120,000 for solar type stars \citep{bouchy_fundamental_2001}. To achieve the science goals for NEID, we use a standard Zerodur 1x2 mosaic R4 Echelle grating with a ruled area that is 200 mm x 800 mm across \citep{schwab_design_2016} in the classic white pupil design \citep{baranne_white_1988}. Using the standard grating equation in Littrow configuration, the resolving power ($R$) is given by 

\begin{equation}\label{eq:resolving power}
    R = \frac{2~d~\rm{tan}(\delta)}{\phi~D},
\end{equation}

where:
\begin{itemize}
    \setlength\itemsep{0em}
    \item $d$ = Beam diameter inside the spectrograph
    \item $D$ = Telescope primary diameter
    \item $\phi$ = On-sky fiber diameter (angular)
    \item $\delta$ = Incident angle of beam on grating  
\end{itemize}

For an R4 Echelle, telescope diameter ($D$) of 3.5 m, beam diameter on the echelle ($d$) of 195 mm, and spectral resolution ($R$) of 100,000, a fiber size of 0.92\arcsec{} balances the need for high spectral resolution and good efficiency by matching the median seeing at WIYN which is $\sim$ 0.8\arcsec{} FWHM median seeing in the R-band (with this fiber diameter, under median seeing conditions only $\sim49\%$ of the light is coupled into the fiber). 

\subsubsection{Fiber Transmission}
\begin{figure}[ht] 
\center
\includegraphics[width=0.95\columnwidth]
{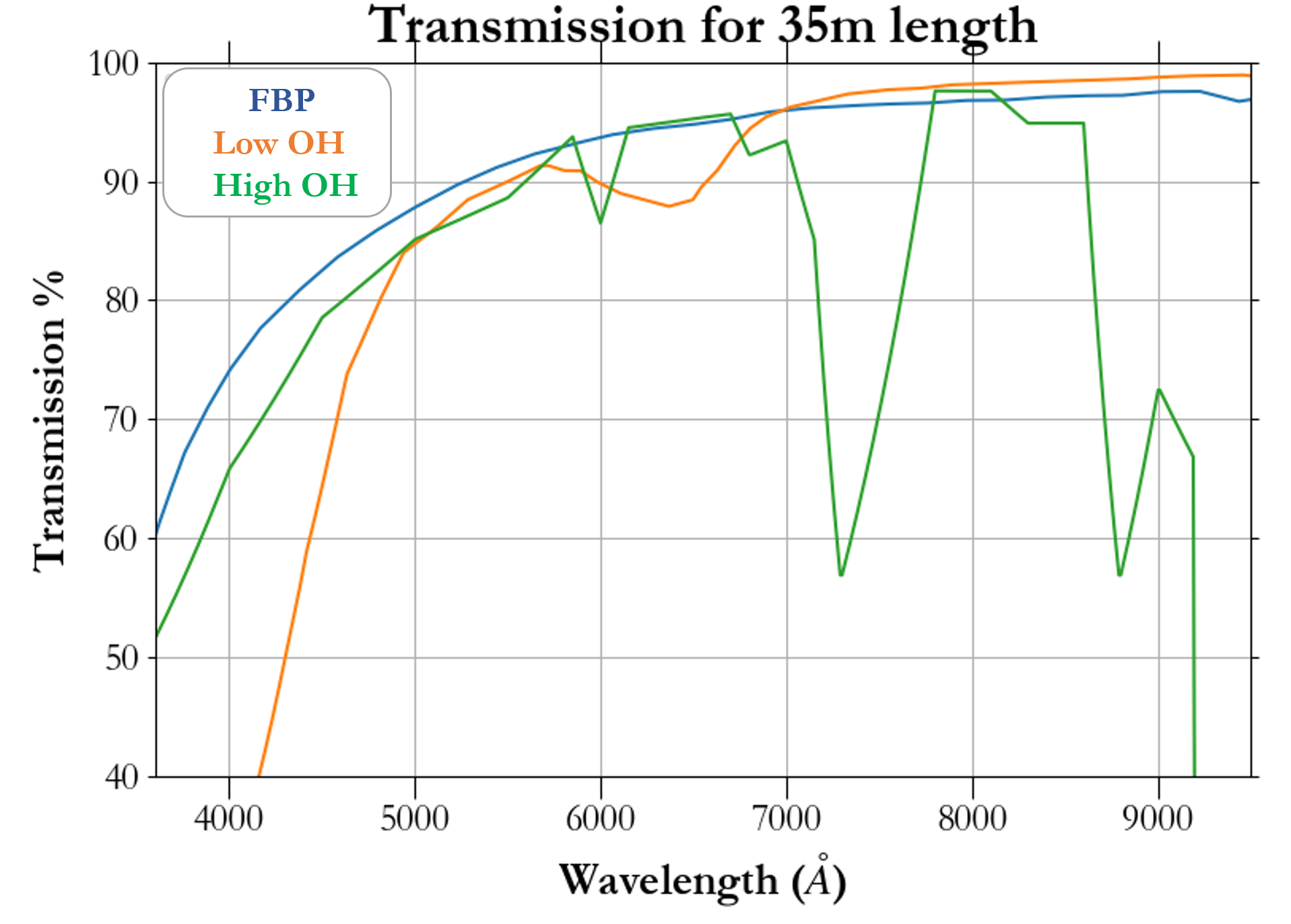}
\caption{Fiber transmission comparison for the FBP fiber used for NEID, versus the low and high OH$^{-}$ variants, for 35 m length (the approximate length of the NEID fiber train).} \label{fig:fibertransmission}
\end{figure}

The photon noise for an RV instrument is inversely proportional to the square root of the number of photons received at the detector, i.e., inversely proportional to the square root of the efficiency of the system.  We estimate that only about $\sim20\%$ of the stellar light is coupled into the instrument fiber after accounting for seeing losses, atmospheric extinction, telescope throughput, etc. Therefore it is imperative to maximize the efficiency of the instrument system, including the fiber feed. NEID uses a single arm spectrometer and fiber with a minimum requirement to cover the entire wavelength bandpass from 380 to 930 nm\footnote{NEID covers 370 to 1090 nm, albeit with the throughput dropping precipitously outside of the nominal bandpass.}, with the total science fiber spanning about 35 m from telescope to instrument. We use the Polymicro FBP substrate for both HR and HE fibers which does not exhibit the OH$^{-}$ ion dip at $\sim 700$ nm like high OH$^{-}$ fibers, while at the same time having lower attenuation in the blue than the low OH$^{-}$ fiber (\autoref{fig:fibertransmission}).

The final photon noise for different stellar \teff{}, magnitudes, and exposure times can be estimated through the NEID Exposure Time Calculator\footnote{\url{http://neid-etc.tuc.noirlab.edu/calc_shell/calculate_rv}}, which is based on the measured on-sky NEID throughput and resolution.

\subsection{Contributing to Instrumental Noise}
\subsubsection{Separate Sky Fiber}
NEID uses a separate Sky fiber to correct for solar contamination from moonlight, atmospheric scattering, and also OH$^-$ emission lines contaminating the science spectrum.  Uncorrected solar contamination can cause systematic RV errors of many \ms{}, the exact magnitude of which depends on the target brightness, sky brightness, zenith angle, lunar phase, etc., \citep{roy_solar_2020}. However, by using a Sky fiber and avoiding observations very close to a bright moon, we allow for an error contribution up to 10 \cms{} from sky subtraction residuals for the NEID instrumental error budget based on results from \citet{roy_solar_2020}.

\subsubsection{Use of Coherent Fiber Bundles}
Three coherent fiber bundles (CFBs) from Schott were originally added to each telescope fiberhead to robustly triangulate the position of the target star relative to the science fiber in the presence of varying telescope flexure. Each CFB consists of 3500 individual single-mode fibers and have a combined size of 6.3\arcsec{} on sky (quality image area of $\sim$ 5.8\arcsec{}). All six CFBs (three from each fiberhead) are imaged on to a single CCD, and have had their positions mapped with respect to the science fiber at the $\mu$m level (see Section \ref{sec:telescope}). While the CFBs are not currently used for target star positioning, CFB images are automatically taken contemporaneously with each nighttime science spectrum to allow an independent estimate of sky brightness.

\subsubsection{Spatial Scrambling}
Achieving the sub \ms{} instrument precision goal for NEID \citep{halverson_comprehensive_2016} necessitates excellent illumination stability from the fiber system. While the simultaneous Calibration fiber can be used to correct for instrumental drift seen by the Science fiber, this assumes that the two fibers see similar changes in RVs due to instrumental effects. This differential fiber-to-fiber correction cannot be used to correct for spurious RV shifts due to relative changes in the illumination pattern of the two fibers, making it important to maintain near-field and far-field illumination stability \footnote{For an introduction to the importance of far-field and near-field stability patterns, refer to \citet{hunter_scrambling_1992}.}. 

Optical fibers already offer extensive azimuthal scrambling, however the scrambling along the radial axis is incomplete \citep{angel_very_1977, avila_photometrical_2006, avila_optical_2008}. This can cause the output intensity distribution to change, as the input illumination varies due to pointing errors, seeing, pupil changes, and atmospheric dispersion. NEID uses a combination of non-circular fibers \citep[octagonal;][]{chazelas_new_2010, avila_frd_2012, spronck_extreme_2012}, and the ball lens double scrambler to maximize scrambling gain \citep{halverson_efficient_2015}, which then dictates the requirements for the port adapter in terms of the pointing (near-field positioning changes) and guiding (far-field chief ray angle variations) systems \citep{logsdon_neid_2018}. This is discussed further in Section \ref{sec:spatialscrambling}.

\subsubsection{Modal scrambling}
Optical fibers used in astronomy are typically multi-mode fibers that allow propagation of a finite number of transverse modes. These modes interfere at the fiber output, giving rise to an interference pattern, commonly referred to as a speckle pattern \citep{hill_modal_1980, rawson_frequency_1980, goodman_statistics_1981}. This pattern depends on the coherence path length of the light source, fiber diameter, and the wavelength of light, and can change as the fiber is perturbed due to telescope movement or temperature fluctuations. These changes manifest in the far-field output of the fiber, which can add to the spectrograph RV noise due to centroid offsets \citep{baudrand_modal_2001, lemke_modal_2011, mahadevan_suppression_2014}. For the instrument fiber, we mitigate this using a mechanical agitator based on the one used in HPF \citep{mccoy_optical_2012, mahadevan_suppression_2014, roy_scrambling_2014, kanodia_overview_2018}. Given that the calibration light from the photonic sources (LFC and etalon) is more coherent than the continuum stellar light, we use a separate mechanical scrambler for the calibration sources, made by GigaConcepts\footnote{\url{http://gigaconcept.com/index.html}}. To empirically test the modal-noise contribution from the calibration source, we obtained a series of LFC observations with NEID alternating the modal-noise agitator (common for Science, Sky and Calibration fibers) on/off. Through these exposures, the mechanical scrambler made by GigaConcepts is kept on as it is part of the standard calibration sequence with the LFC and etalon. We take four sequences of data, where each one is defined as follows - 

% NEID uses a custom LFC developed by Menlo Systems, who also built and tested the LFC for HARPS.

% Based on fiber-to-fiber tracking tests from the HARPS LFC, we put an upper limit of 2.0 \cms{} on the modal noise contribution for both the Science and Calibration fibers.

% Given the coherent nature of the LFC and etalon flux,

\begin{itemize}
    \item Agitator On: 6x 60 s \vspace{-0.2cm}
    \item Agitator Off: 3x 60 s \vspace{-0.2cm}
    \item Agitator Movement \vspace{-0.2cm}
    \item Agitator Off: 3x 60 s \vspace{-0.2cm}   
    \item Agitator Movement \vspace{-0.2cm}
    \item Agitator Off: 3x 60 s \vspace{-0.2cm}
    \item Agitator On: 6x 60 s \vspace{-0.2cm}   
\end{itemize}

\begin{figure*}[!t]
    \begin{minipage}[b]{0.5\linewidth}
    \centering
    \includegraphics[width=0.9\textwidth]{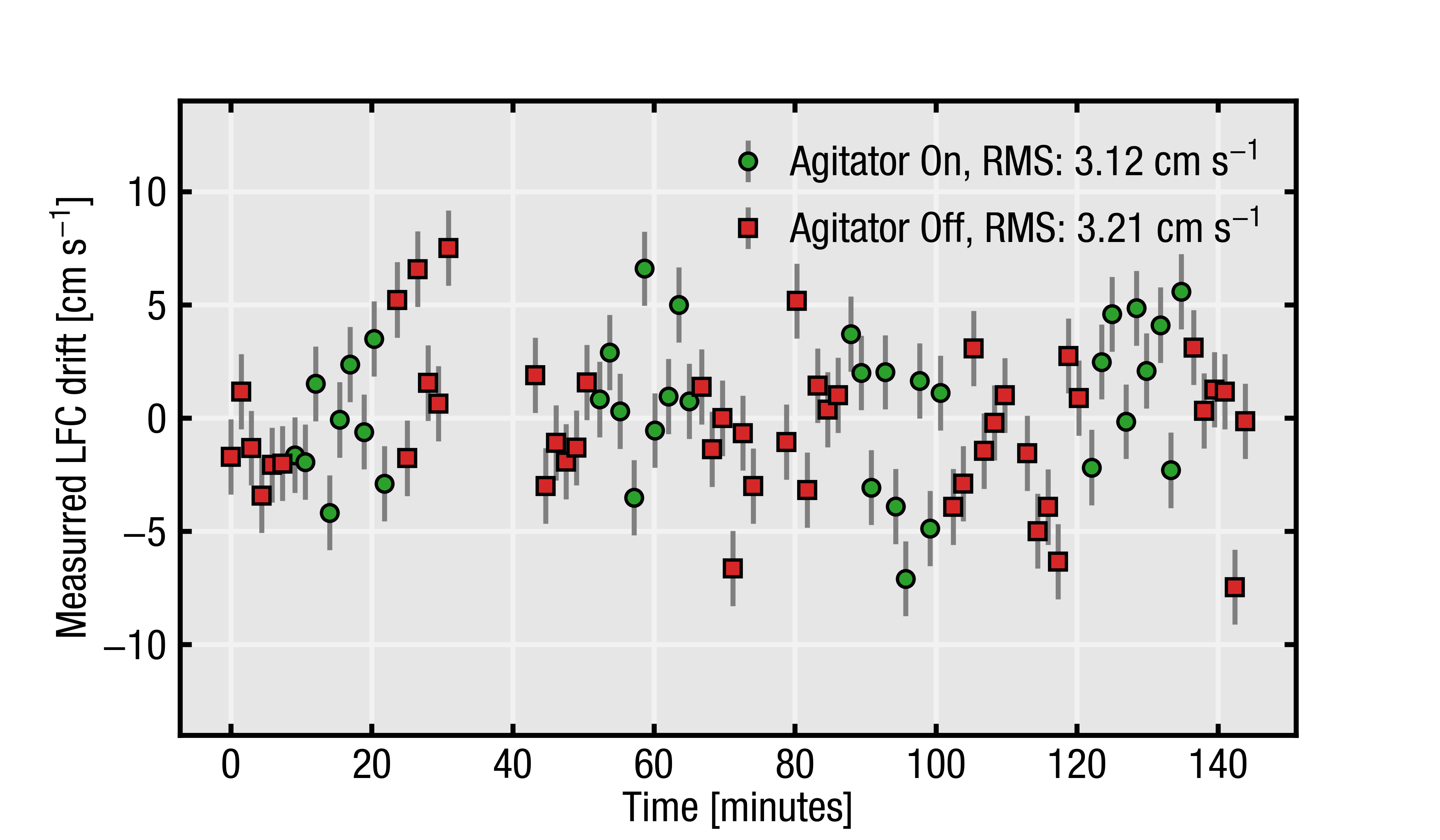}
    \end{minipage}
    \begin{minipage}[b]{0.5\linewidth}
    \centering
    \includegraphics[width=0.9\textwidth]{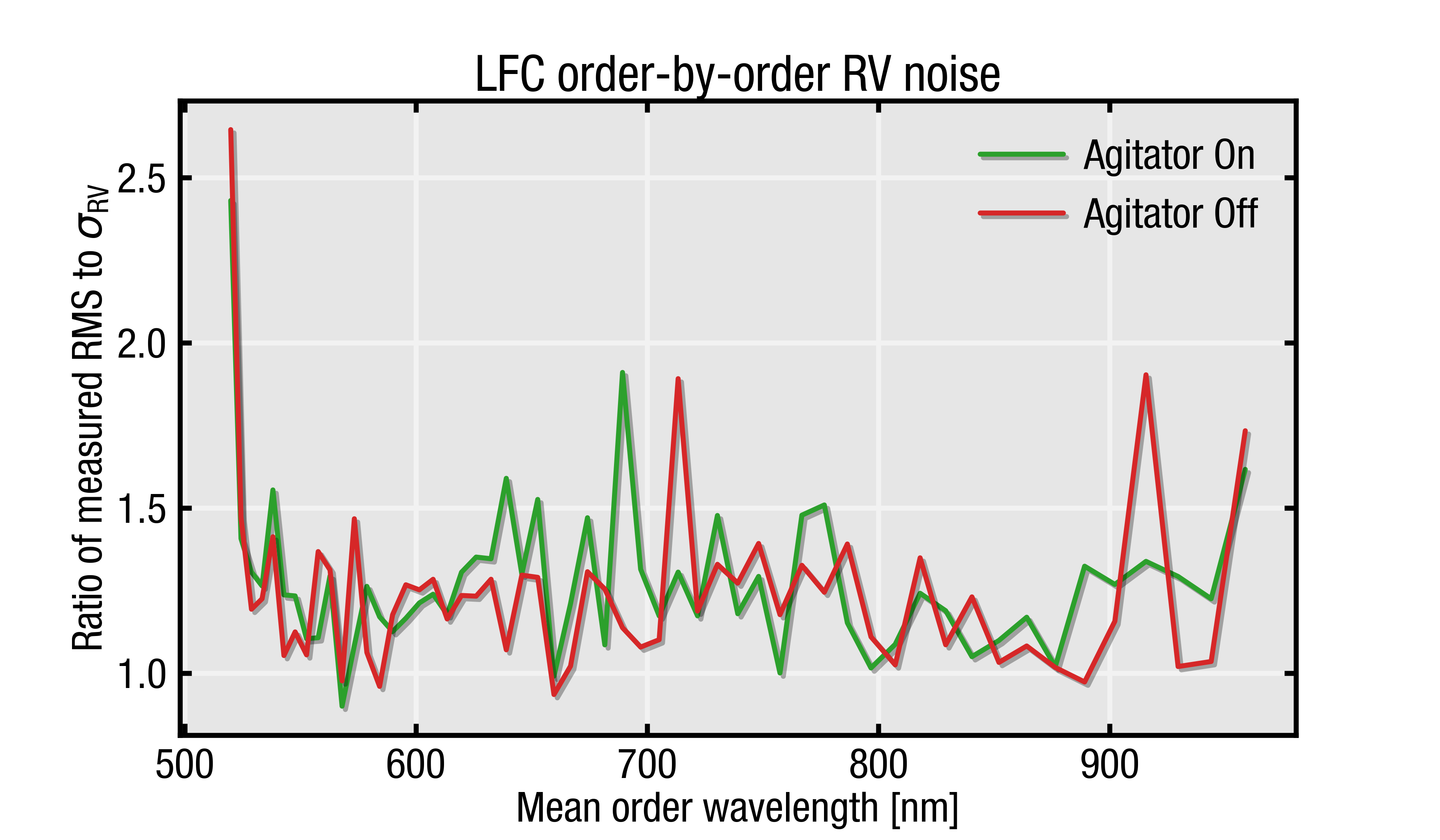}
    \end{minipage}
  \caption{(a) We show the LFC time-series with the modal-noise agitator on and off sequences interspersed by agitator movement to mix the fiber modes. Given our finite sample size and photon noise, we consider the RMS difference between the two agitator cases --- on (green circles) and off (red squares) to be similar. b) We compare the ratio of RMS scatter to the photon noise across each order index (0 is for the bluer wavelengths, and 55 towards the redder wavelengths), and do not see any chromatic trends suggestive of the $\lambda^2$ dependence of modal-noise.}   \label{fig:ModalNoise}
\end{figure*}

The agitator movement comprises of turning it on, waiting for 10 seconds, and turning it off. This is meant to move the fibers, thereby changing the modal distribution within the fibers, where during the 'Off' exposures, the agitator is kept static. This entire sequence is repeated four times resulting in a total of 84 LFC exposures. In \autoref{fig:ModalNoise}a we see that the RMS scatter is comparable across the two agitator states, which suggests that the GigaConcepts scrambler is performing most of the scrambling of the calibration light. We then compare this RMS scatter with the estimated photon noise for the LFC spectrum and find the RMS to be $\sim$ 1.3x the photon noise across most of the NEID orders which receive LFC flux. Given the estimated photon noise of 2.5 \cms{} and an RMS of $\sim 3.2$ \cms{}, we ascribe this additional noise component of $\sim$ 2 \cms{} to be the upper limit on the modal noise term for the calibration source to the NEID error budget.

\subsubsection{Focal Ratio Degradation and fiber stresses} \label{sec:frd_description}

For astronomical spectrographs, both the throughput and the focal ratio degradation (FRD) play a critical role in the performance of fiber-fed instruments. FRD is the tendency of the fiber output to be at a lower f/$\#$ (``faster'')\footnote{f/$\#$ = Focal Length / Diameter of beam} than the input \citep{ramsey_focal_1988}. The HR mode fibers used in NEID have a relatively small core of $\sim 60~\mu$m, which is driven by the fiber size considerations detailed in Section \ref{sec:fiber_size}, and small core fibers tend to suffer more from the effects of FRD \citep{ramsey_focal_1988}. Unmitigated FRD can increase the output cone angle of the fibers, which can not only lead to a reduction in throughput, but also cause an increase in the scattered light background on the detector \citep{kanodia_ghosts_2020}. Therefore, throughout the design and fabrication process for the NEID fiber train, utmost care was taken to minimize any additional sources of FRD for the fibers in the system, including the vacuum feedthrough where the fibers pass into the vacuum chamber.  The two largest sources of FRD are typically stress on the fiber at the terminations as well as small bend radii. We prepared six sets of spare fiber bundles, which were each tested after every fabrication step for throughput and FRD to ensure consistency, and the best performing fiber bundle was installed in NEID.  We discuss the testing procedure in Section \ref{sec:testing}.

\begin{figure*}[!ht] 
\center
\includegraphics[width=\textwidth]
{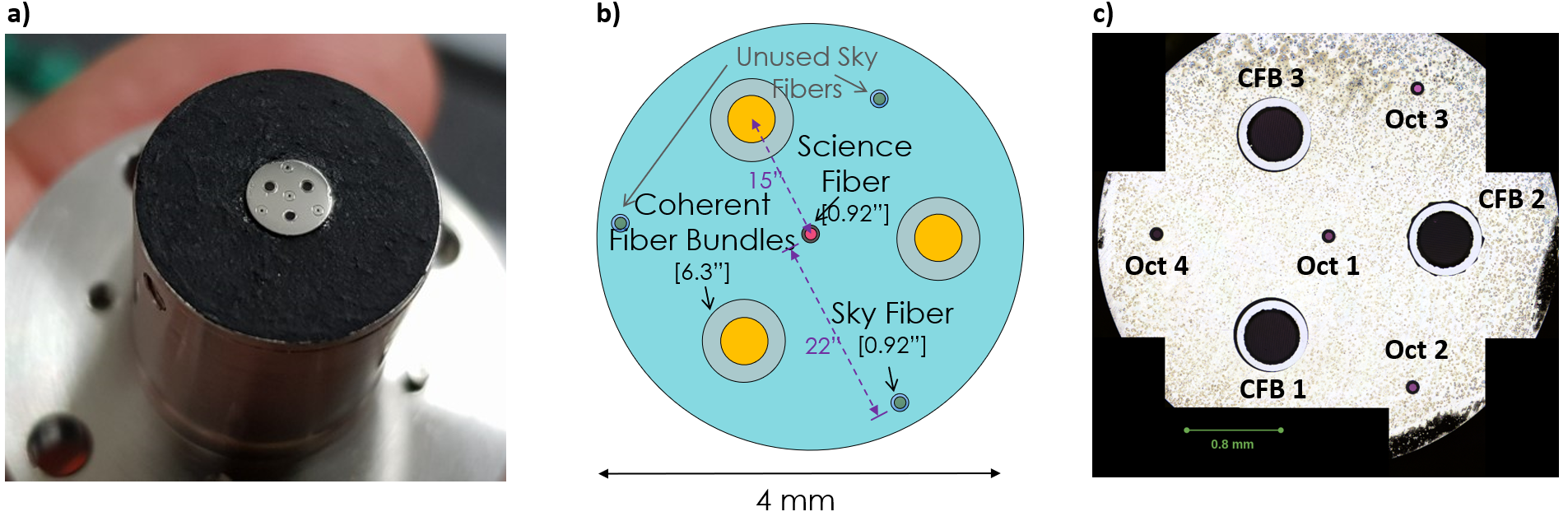}
\caption{\small The NEID telescope fiberhead. \textbf{a)} Image showing the fiberhead, with a black coating to reduce backscatter. \textbf{b)} Schematic of the fiberhead adapted from \cite{roy_solar_2020} showing the relative positions of the science and sky fiber with respect to the CFBs. \textbf{c)} High resolution image taken of the fiberhead installed in NEID, taken using a Zygo Optical profilometer. Oct 1 is the Science fiber and Oct 4 is the Sky fiber, while the other octagonal fibers are spares.} \label{fig:fiberhead}
\end{figure*}

\section{Telescope fiber}\label{sec:telescope}

The telescope fiber bundles are 26 m in length and consist of four octagonal fibers (referred to as Oct 1 -- Oct 4) and three CFBs (1 m in length) for the HR mode, and similar for the HE mode. The central octagonal fiber is the Science fiber, on which the star is positioned, while the other three are sky fibers, of which one is used and the other two are spares. The CFBs in the fiberhead (shown in  \autoref{fig:fiberhead}) are 120$^{\circ}$ apart, and feed into the CFB camera in the port adapter \citep{logsdon_neid_2018}. The 120$^{\circ}$ orientation for the fibers and CFBs was chosen to avoid diffraction spikes (at 45$^{\circ}$) from the science fiber contaminating the sky fibers or the CFBs.  We had a total of five telescope fiber bundles fabricated by Berlin Fibre\footnote{\url{https://www.berlin-fibre.de/en/}}, which were numbered sequentially and included one test bundle, and two each for the HR and HE modes. This allowed for one spare fiber bundle for each mode. The final units installed in the port adapter were chosen based on the throughput and FRD testing performed at Penn State and listed in \autoref{tab:throughputfrd}.

\iffalse
\begin{figure*}[!t]
  \centering
  \begin{tabular}[b]{@{}p{0.4\textwidth}@{}}
  {
    \centering\includegraphics[width=8cm]
{fiberhead.jpg}} \\
\centering\small (a) NEID telescope fiberhead showing the three CFBs and four octagonal fibers .
  \end{tabular}%
  \quad
  \begin{tabular}[b]{@{}p{0.4\textwidth}@{}}
 {
    \centering\includegraphics[width=7cm]
{fiberhead_SW.jpg}} \\
    \centering\small (b) Drawing showing the fiberhead and the orientation \citep{logsdon_neid_2018}.
  \end{tabular}
  \caption{SolidWorks drawing of the fiberhead showing the three CFBs and four octagonal fibers. Also shown is the position of the CFBs.}   \label{fig:fiberhead}
\end{figure*}
\fi

\begin{figure*}[!t]
    \begin{minipage}[b]{0.5\linewidth}
    \centering
    \includegraphics[width=0.6\textwidth]{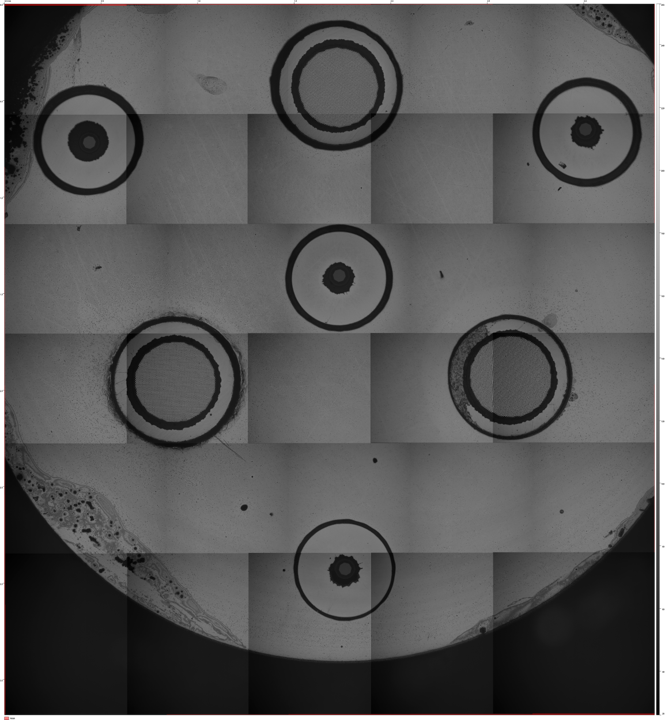}
    \end{minipage}
    \begin{minipage}[b]{0.5\linewidth}
    \centering
    \includegraphics[width=0.8\textwidth]{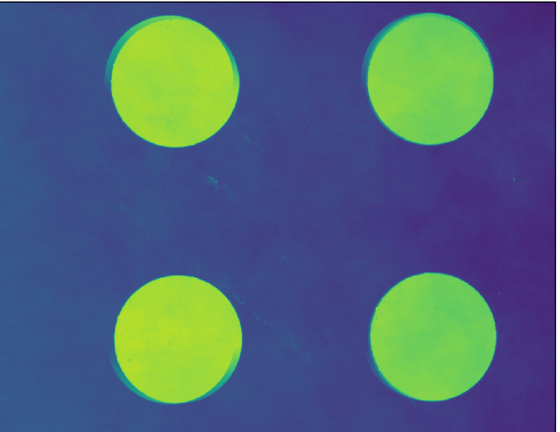}
    \end{minipage}
  \caption{(a) Stitched collection of individual frames using the weighted averages algorithm, showing the stitch boundaries. This was later replaced by the `High Contrast' algorithm to stitch the images together. The fiberhead is 4 mm in diameter. (b) Thorlabs polka dot pattern showing the offsets due to the imperfect stitching. Each dot is 125 $\mu$m in diameter.}   \label{fig:zygo_stitching}
\end{figure*}

% \subsection{Precise metrology of fiberhead}\label{sec:metrology}
The NEID error budget \citep{halverson_comprehensive_2016} dictated the instrument requirements for each component of the telescope, port adapter and spectrograph. As discussed by \cite{logsdon_neid_2018}, the scrambling gain for the NEID fiber-feed translates into a stellar image positioning tolerance of $\sim 2.7 \mu$m, which necessitates precise metrology of the NEID fiberhead giving the position of the each fiber and CFB at the 1 $\mu$m level. We performed this metrology in the lab using a telecentric lens as follows:
\begin{enumerate}
\item{Calibrated the plate scale and radial distortion profile of the telecentric lens using a Polka dot pattern\footnote{Fixed Frequency Grid Distortion Targets with $250 \pm 1 \mu$m spacing, Thorlabs R2L2S3P2}.}
\item{Imaged the fiberhead to obtain the positions of the fibers and CFBs in pixel space, then corrected them using the distortion estimates, and converted to physical units (mm) using the plate scale measurements from above.}
\end{enumerate}
The telecentric lens measurements were then verified using a Zygo Optical Profilometer. The profilometer relies on white light interferometry to measure the 3-D profile of a surface, and also a very precise 2-D map of the fiber locations. The steps followed were as follows:
\begin{enumerate}
    \item{Calibrated the plate scale of the imager using a UV grating with a known spacing of 3.455 $\mu$m, measured using an atomic force microscope (AFM). We measured the grating pitch of the imager by fitting a peak to the image in 2D Fourier space, and matching that to the AFM measurements. With a 20x magnification, and 0.5x zoom, we obtained a plate scale of 0.8183 $\mu$m/pixel for the 1024 x 1024 pixel grid.}
    \item{Imaged the telescope fiberhead, while simultaneously back-illuminating the fibers.}
\end{enumerate}

The profilometer uses white light optical profilometry \citep{groot_principles_2015} along with piezo motors to obtain nm level precision in the \textit{z} (depth) axis. However, it is also a powerful tool for precisely determining spatial distances in the \textit{x} - \textit{y} plane to get the relative positions of the fibers. We used the intensity mode to measure the spatial positioning of the fibers, in which the profilometer is essentially used as an ordinary microscope with a 20x magnification and 0.5x zoom.

It was challenging to distinguish the core and the cladding of the octagonal fibers when centroiding them in the 3D mode, since there is no surface feature distinguishing the two (similar topography). We therefore used the intensity mode by back-illuminating the octagonal fibers to increase the contrast between core and cladding. The high magnification necessitated stitching of multiple individual frames in order to obtain a precise scan of the entire fiberhead (\autoref{fig:zygo_stitching}a) which was accomplished using an automated precise x-y stage. This stitching process had an error of the order of 2 -- 3 pixels (about 2 $\mu$m) which exceeded the tolerances on this measurement (\autoref{fig:zygo_stitching}b). We attempted to diagnose the magnitude of the stitching errors using a 250 $\mu$m polka dot beam-splitter grid from Thorlabs. \autoref{fig:zygo_stitching}b shows the polka dot pattern with the offsets induced by the inaccurate stitching. This issue was diagnosed in consultation with the manufacturer by switching to the higher contrast (instead of the weighted average) algorithm for removing fringes. This improved fringe removal allowed for better stitching of the individual frames.

Furthermore, to increase the contrast between the octagonal core and corresponding cladding in the intensity images we used a `True Color' mode \citep{beverage_interferometric_2014}. This mode is essentially a color image of the surface using multi-color illumination, which is then superimposed on the 3D phase map. The polka dot stitching was repeated with this configuration, following which the actual fiberhead was remeasured and is shown in \autoref{fig:fiberhead}c. This finally enabled accurate measurement of the distances between the fibers and CFBs. 

The measured distance between fibers was converted to physical units using the plate scale we estimated using the UV grating. The new plate scale, coupled with the improved stitching gave us results comparable to the telecentric lens measurements ($< 1 \mu$m), thereby verifying they met our metrology requirement. The measured coordinates for the octagonal fibers and CFBs for HR2 and HE5 (the installed fiber bundles) are given in \autoref{tab:TelescopeHR} and \autoref{tab:TelescopeHE} in the Appendix, and have a typical uncertainty of 0.1 $\mu$m for the positions of the octagonal fibers, and 0.8 $\mu$m for the CFBs. 
Lastly, we measured the Focal Ratio Degradation (FRD; Section \ref{sec:frd}) and throughput (Section \ref{sec:throughput}) for the octagonal fibers in the telescope fiberhead.

% We then measured the relative orientation of the fibers between the input and output end of the CFB. This was done using a shadow from an obstruction introduced on the output end while  the input end was imaged using a point source microscope. Since the shadow was deterministically oriented with respect to the fiberhead, the relative orientation of the output can be determined (\autoref{ch3:fig:cfb_shadow}).

\section{Spatial Scrambling}\label{sec:spatialscrambling}

\begin{figure}[!t] 
\center
\includegraphics[width=0.4\textwidth]
{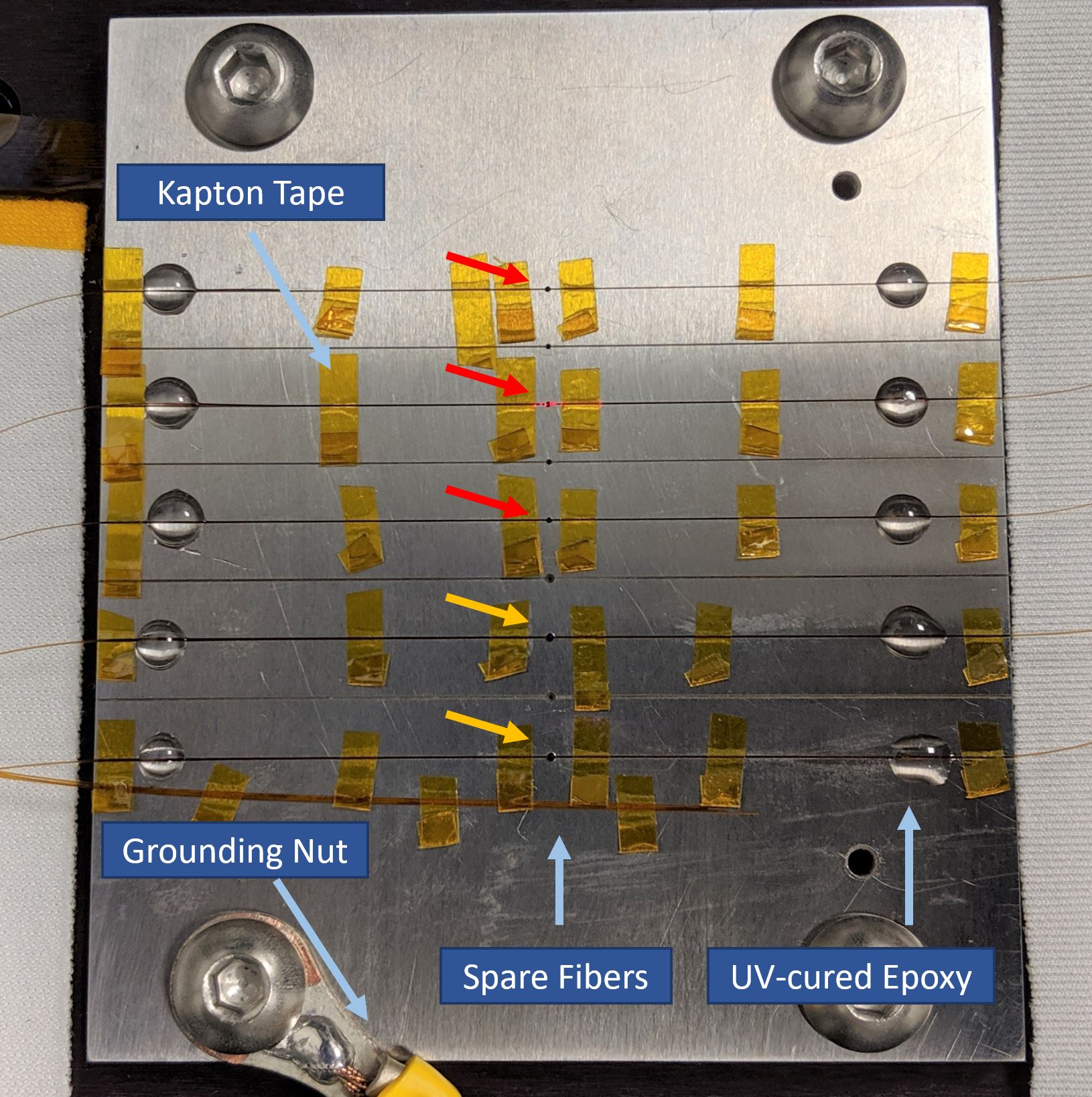}
\caption{The NEID v-groove block with the 5 fibers (3 HR + 2 HE) in the process of alignment to the ball lenses during commissioning at WIYN. The yellow arrows point to the two HE fibers and ball lenses, while the red ones indicate the three HR fibers and lenses, with one of them lit up with an alignment laser. We ground the stainless steel block to prevent electrostatic discharges from disturbing the tiny lenses, and used Kapton tape and epoxy to fix the fibers in place upon alignment. Lastly, this block is then capped with a foam-covered lid that is not shown in this image.} \label{fig:vgroove}
\end{figure}

\begin{figure*}[!t]
    \begin{minipage}[b]{0.5\linewidth}
    \centering
    \includegraphics[width=0.6\textwidth]{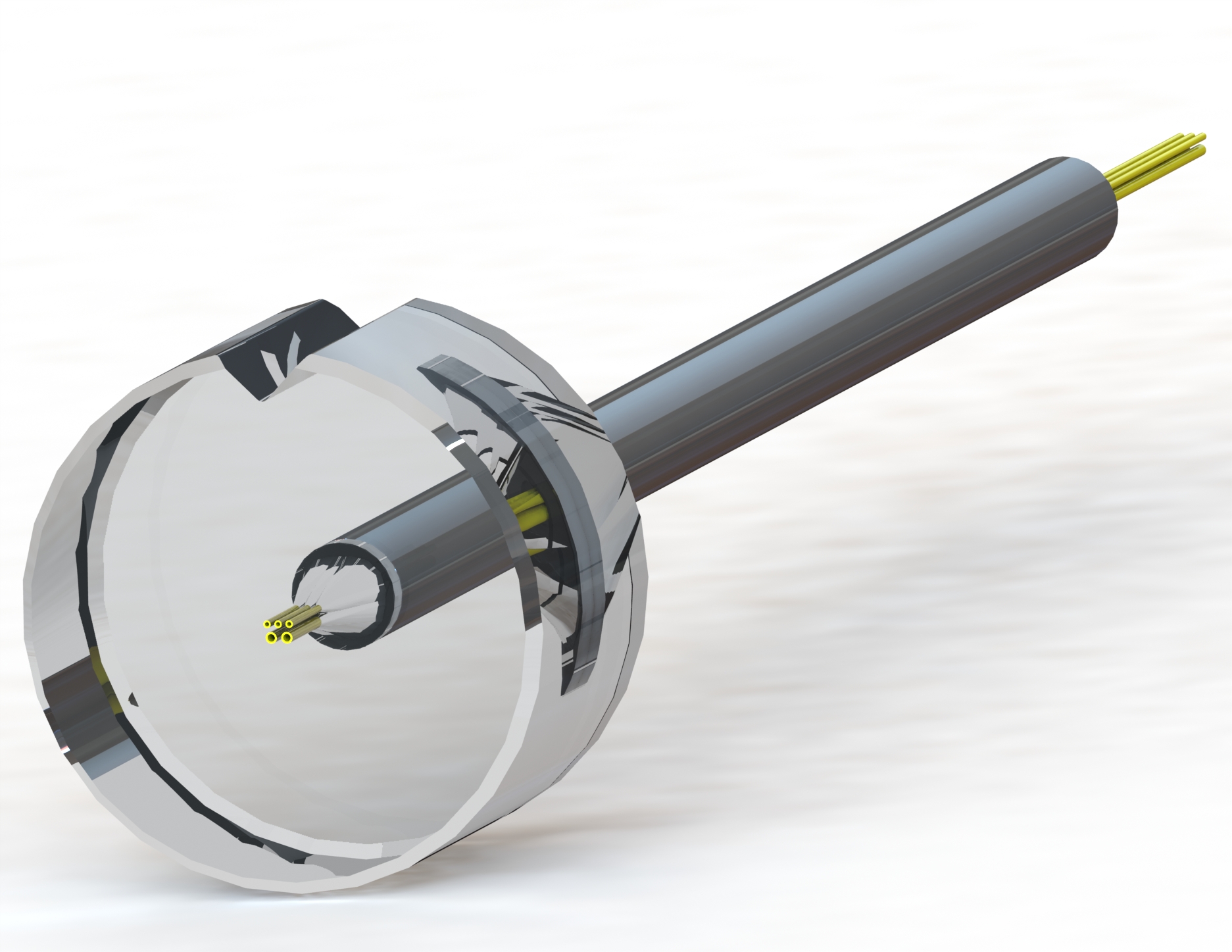}
    \end{minipage}
    \begin{minipage}[b]{0.5\linewidth}
    \centering
    \includegraphics[width=0.8\textwidth]{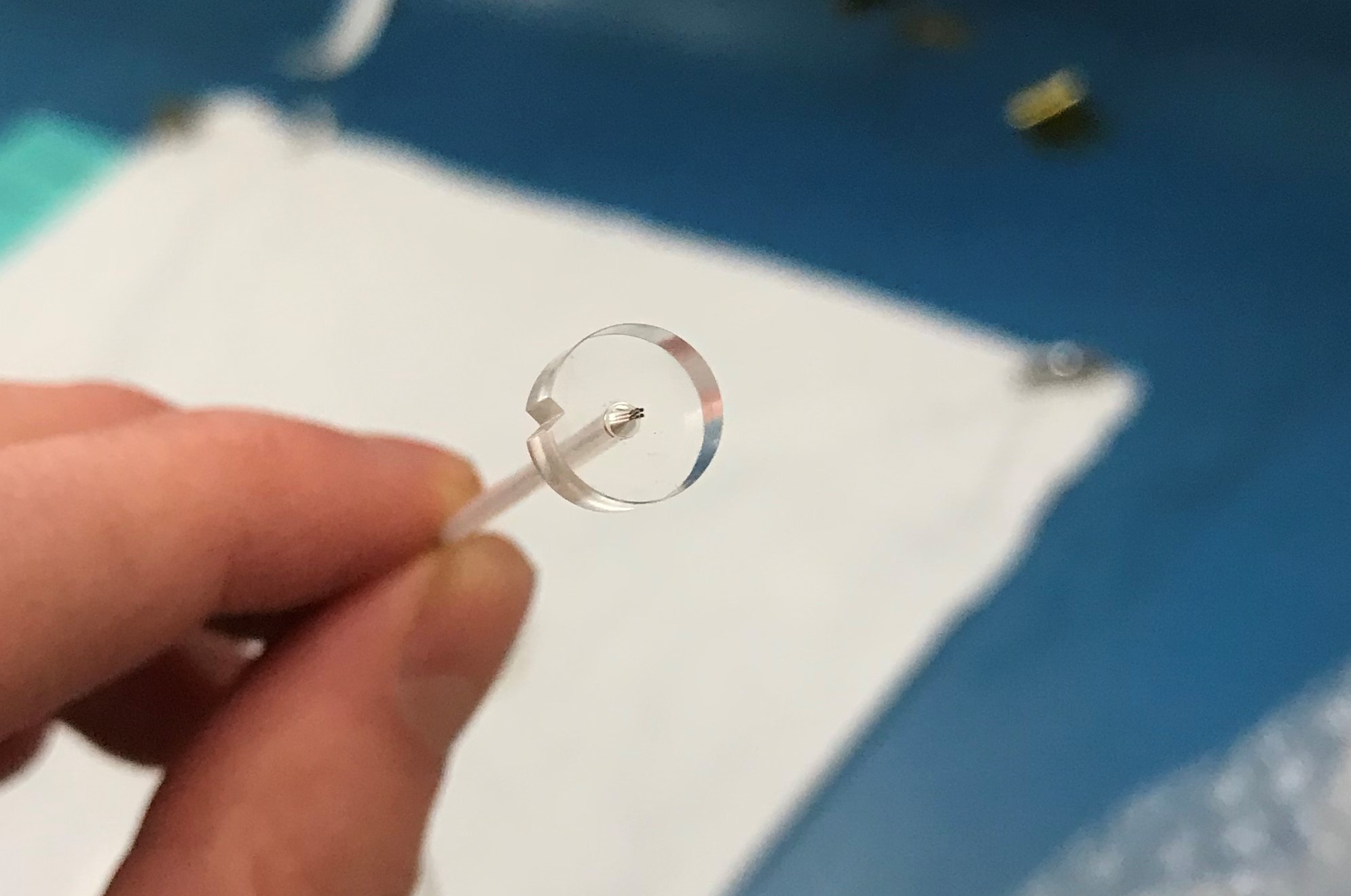}
    \end{minipage}
  \caption{The NEID fused-silica puck effectively functions as the NEID slit with three HR fibers (Calibration, Science and Sky) and two HE fibers (Science and Sky). (a) SolidWorks rendering of the puck showing the five fibers, and (b) the NEID puck after potting the fibers and polishing to the final finish.}   \label{fig:puck}
\end{figure*}

Excellent image and pupil scrambling is required to ``desensitize'' the instrument from external variations in input illumination. The extent to which this is done is characterized as the scrambling gain of the system, which governs the requirements on the various sub-systems of the port adapter.  For NEID we use the same ball lens double scrambler concept as for HPF, albeit with a much smaller ball lens due to the smaller fiber diameter.  The HR ball lens is 500 $\mu$m in diameter, and is made of Hoya glass TaFD-45 with a refractive index $n \sim 2$. Chromatic deviations from $n \sim 2$ result in a degradation in throughput and scrambling performance which are encapsulated in the total NEID throughput estimate after scaling by throughput measurements at 630 nm (Section \ref{sec:throughput}).

The inputs to the ball scrambler are the octagonal HR and HE Science and Sky fibers from the telescope fiber bundles, along with an octagonal HR Calibration fiber from the calibration bench (\autoref{fig:vgroove}). These are face-coupled to the ball lens in a precise EDM (Electrical Discharge Machined) stainless steel v-groove block. At the output end, we have a 2 m section of octagonal fiber face-coupled to the ball lens on one end, with this end of the octagonal fiber also coated with a broadband AR coating to maximize throughput, and spliced to the circular instrument fiber on the other. We use a fusion splicer\footnote{FITEL S183PM II Fusion Splicer} to fuse the two fibers, and follow the same methodology as prescribed for HPF \citep{kanodia_overview_2018}, while optimizing the splicer fusion parameters to give the best FRD and throughput performance for the NEID fiber feed. This combination of an octagonal fiber, double scrambler, and an octagonal fiber spliced to a circular fiber has been tested to maximize scrambling gain \citep[Table 1 in ][]{halverson_efficient_2015}. The same spatial scrambling system is used in HPF, where the far-field scrambling performance has been validated on-sky \citep{kanodia_harsh_2021}.

\section{Instrument Fiber}\label{sec:instrument}

Inside the NEID cryostat, the five fibers (3 HR and 2 HE) in the fiber bundle are potted inside a fused silica puck (\autoref{fig:puck}). This puck was manufactured by Femtoprint\footnote{\url{https://www.femtoprint.ch/}} using the same fabrication procedure as the HPF puck, which achieves the tight tolerances for the fiber bore centroids \citep[$< 1 \mu$m;][]{kanodia_overview_2018}. The NEID optical fiber core and cladding are made from fused silica, so the matching coefficient of thermal expansion \edit1{of the puck} reduces stresses from compression of the puck as the instrument goes from room temperature to its operating temperature of $\sim$ 300 K. The puck has five bore-holes sized to the NEID fibers and conical openings leading to each bore-hold to help ease the fibers into the bores.

The fibers were glued in the bores using Epotek 301-2 epoxy and allowed to cure at room temperature over 48 hours. After this, we polished the fibers by mounting the puck in an aluminium mount on the polishing table, and then polishing the fibers using the method followed for HPF \citep{kanodia_overview_2018}.

Once the puck was polished and tested, we glued a fused silica singlet lens on the puck using Epotek 301-2 epoxy. This is the first lens in the set of lenses in the optics tube, which increases the f-ratio from f/3.65 to f/8.1 and also provide a pupil stop that limits stray light. The design, optimization and implementation for this optical configuration are further discussed in \cite{schwab_neid_2020}.

A v-shaped wedge is cut into the puck to help clock and lock it in place inside its mount. This is important to ensure the correct alignment of the 3 fibers on the NEID detector, after correcting for the field rotation due to the $\gamma$ angle from the echelle grating \citep{hearnshaw_astronomical_2009}.

\section{Testing procedure}\label{sec:testing}

\begin{figure*}[!htbp]
    \begin{minipage}[b]{0.5\linewidth}
    \centering
    \includegraphics[width=\textwidth]{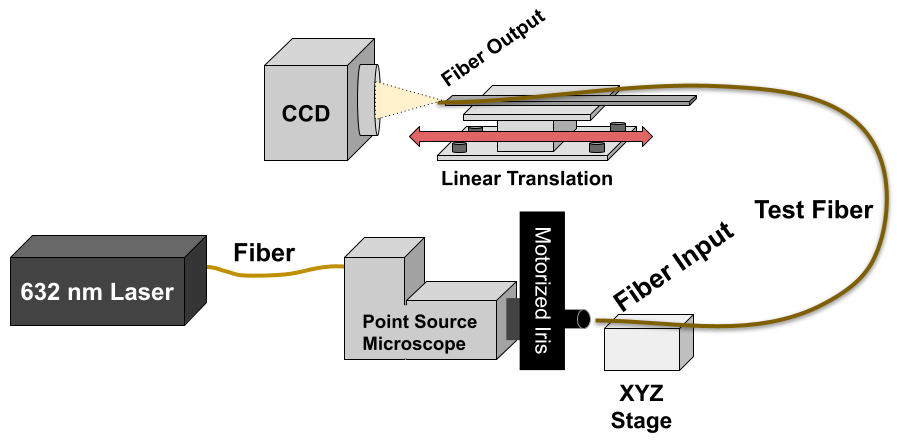}
    \end{minipage}
    \begin{minipage}[b]{0.5\linewidth}
    \centering
    \includegraphics[width=\textwidth]{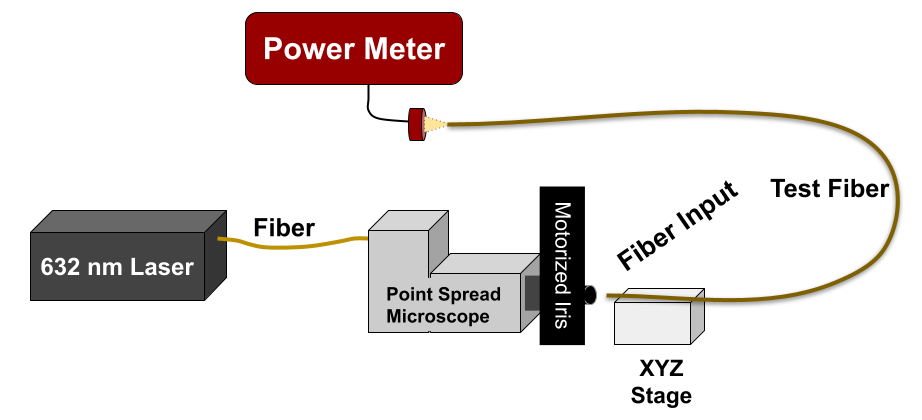}
    \end{minipage}
  \caption{Block diagram showing the setup used to measure the (a) FRD and (b) throughput for the test fiber. We use a point source microscope to ensure that the test fiber face is normal to the principal axis of the input beam, where the input light (Thorlabs TED4015 with a 635 nm pig-tailed SMF fiber output with $\sim$ 10 nm bandwidth) is focused on the fiber at a deterministic f/\# using a motorized iris. The beam is centered on the fiber face by translating the face using an XYZ stage. The throughput was performed as a differential measurement using a power meter.}   \label{fig:blockdiagram}
\end{figure*}

\subsection{Focal Ratio Degradation}\label{sec:frd}

As mentioned in Section \ref{sec:frd_description}, FRD in the system is an indicator of stresses on the fiber, typically near the terminated ends \citep{ramsey_focal_1988}. Excessive FRD overfills the white pupil on the grating, which degrades the efficiency of the system and also contributes to scattered light inside the instrument, which can present problems in background subtraction and PSF estimation \citep{kanodia_ghosts_2020}.
The procedure to estimate the FRD of each fiber in the NEID system was the same as for HPF \citep{kanodia_overview_2018} and is shown in \autoref{fig:blockdiagram}a. A point source microscope along with a motorized iris couples light into the fiber at a f/4 input. We use a motorized linear stage to project the output cone from the fiber onto a CCD. We then measure the image size as a function of stage translation to estimate the output f/$\#$ of the system. The final measurements are included in \autoref{tab:throughputfrd}, where the FRD measurements for the HR Sci fiber suggest that the enclosed energy at the expected f/out is $\sim$ 90\%.

\subsection{Throughput}\label{sec:throughput}

To measure the throughput of the NEID fibers in a repeatable manner, we used the same input configuration as for the FRD test, with a deterministic input f/$\#$. At the output end we used a Thorlabs photometer to measure the efficiency of the system. We used a Thorlabs diode laser to provide a stable input light source, which we calibrated both before and after the tests, to ensure consistency. The final measurements for the fibers are included in \autoref{tab:throughputfrd}, where the throughput for the HR Sci fiber is about 98\% of the theoretical expected value after accounting for Fresnel and reflection coating losses. We note that these throughput measurements are agnostic of the angular distribution of the output flux, and therefore to get an effective throughput relevant to the acceptance f/$\#$ of the instrument and baffling, the throughput estimates should be multiplied by the FRD.

%Focal Ratio Degradation (FRD) \citep{ramsey_focal_1988}. Minimum stress, slow cure epoxy, minimum bend radius, gentle terminations, match F. Silica puck.. small core fibers have worse FRD. Low shrinkage epoxy - cures and shrinks, introducing microbends. FRD vs SCrambling

\section{On Sky Tests}\label{sec:onsky}
During the NEID instrument commissioning period in Fall 2020/Spring 2021, we performed a variety of stress-tests on the spectrometer to verify the image and pupil scrambling performance of the delivered NEID science fiber system. These tests used both stellar targets and internal calibration sources (LFC, Etalon) to constrain the scrambling performance of the NEID fibers. 

\subsection{Near-field tests}
To probe sensitivity to input near-field variations (e.g., guiding offsets), we used the NEID fiber injection port calibration fiber, which delivers light from the NEID calibration bench in the WIYN basement to the port, to record calibration spectra taken at controlled alignment offsets. The image of the port calibration fiber ($\sim$ 100 ${\mu}$m core diameter) was systematically translated across the NEID science fiber ($\sim$ 62 ${\mu}$m diameter) in discrete steps using the \textit{x}-\textit{y} stage the port calibration fiber is mounted on. A series of etalon spectra was recorded at each spatial step, and the relative velocity offsets were computed. Exposure times were adjusted accordingly to maintain constant spectral signal-to-noise in the etalon spectrum. This was required to remove potential systematics due to charge transfer inefficiency. 

\begin{figure}[t] 
\center
\includegraphics[width=\columnwidth]{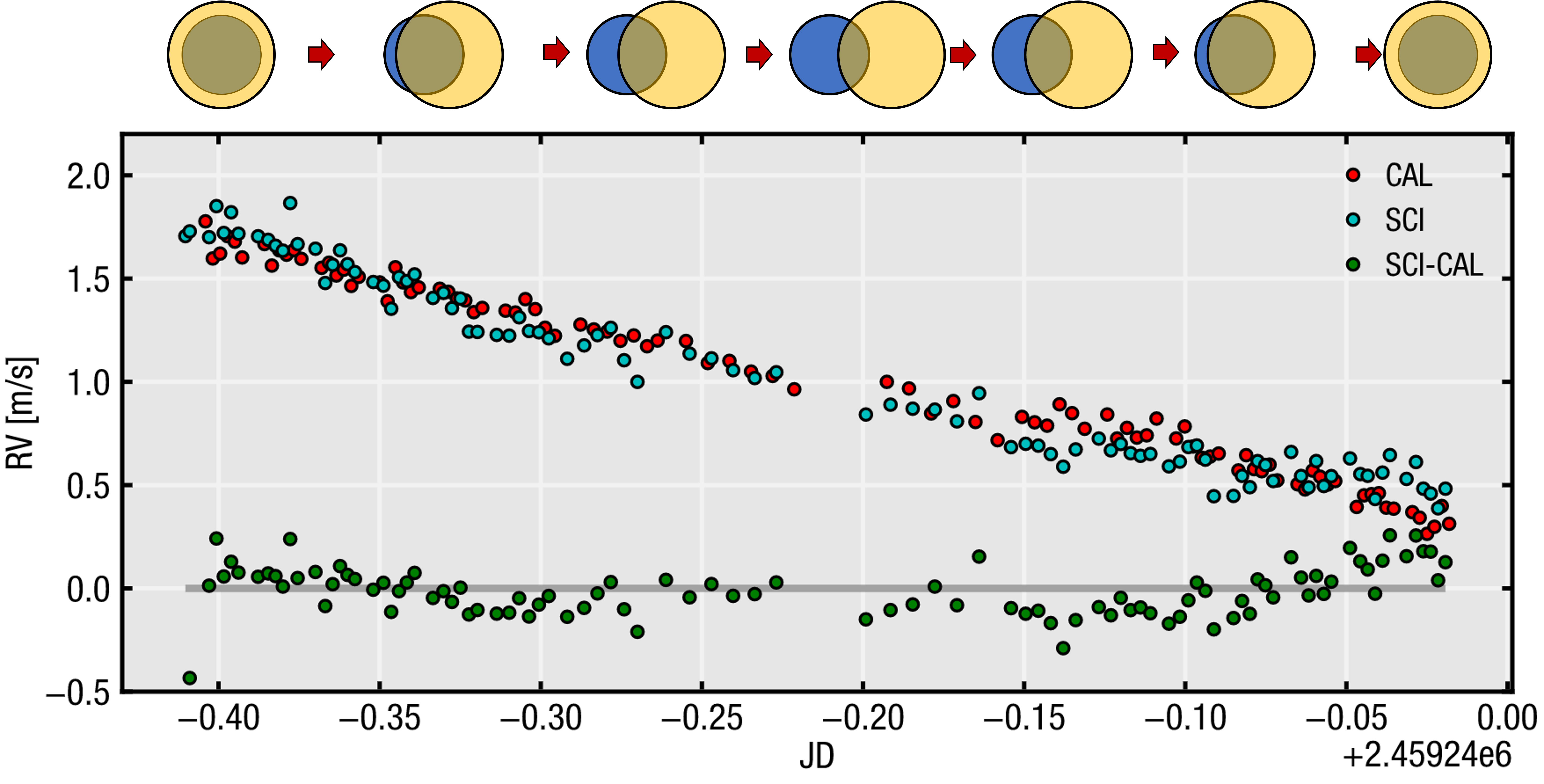}
\caption{Near field scrambling: Measured differential drift between the NEID Science (SCI) and Calibration (CAL) fibers during the near-field scrambling test using the telescope fiber port. The cartoon above shows the sequence of offsets applied to the port calibration fiber (yellow) as the image is stepped across the Science fiber (blue). The NEID etalon calibration source was used to alternately illuminate both fibers during this test. Upon subtracting the intrinsic instrumental drift (measured using the Calibration fiber, red points), the residual Science fiber drifts (green points) are consistent with a near-field scrambling gain of over 10,000.}
\label{fig:scrambling_near-field}
\end{figure}

To separately track instrumental drift, exposures of calibration light through the instrument Science fiber were interleaved with standard calibration frames where only the instrument Calibration fiber was illuminated with the etalon source. This \lq{}nodding\rq{} between Science and Calibration fibers allowed for a precise tracking of the bulk instrument drift during the measurement period, and also ensured both spectra were not affected by cross-contamination between channels (as would have been the case if we had illuminated both the Science and Calibration fibers simultaneously). Figure~\ref{fig:scrambling_near-field} shows the measured relative drifts between the Science (SCI) and Calibration (CAL) fibers during this test period. The image of the port calibration fiber was stepped across the Science fiber in discrete steps of $\sim$20 microns, starting on-center, and extending out to 70 microns off center. This stepping sequence was then reversed, and the resultant differential drift time series (SCI - CAL, green points) shows systematic errors consistent with a near-field scrambling gain of $>$10,000, consistent with laboratory measurements of the analogous HPF fiber system \citep{halverson_efficient_2015}.

\subsection{Far-field tests}

\begin{figure}[t] 
\center
\includegraphics[width=\columnwidth]{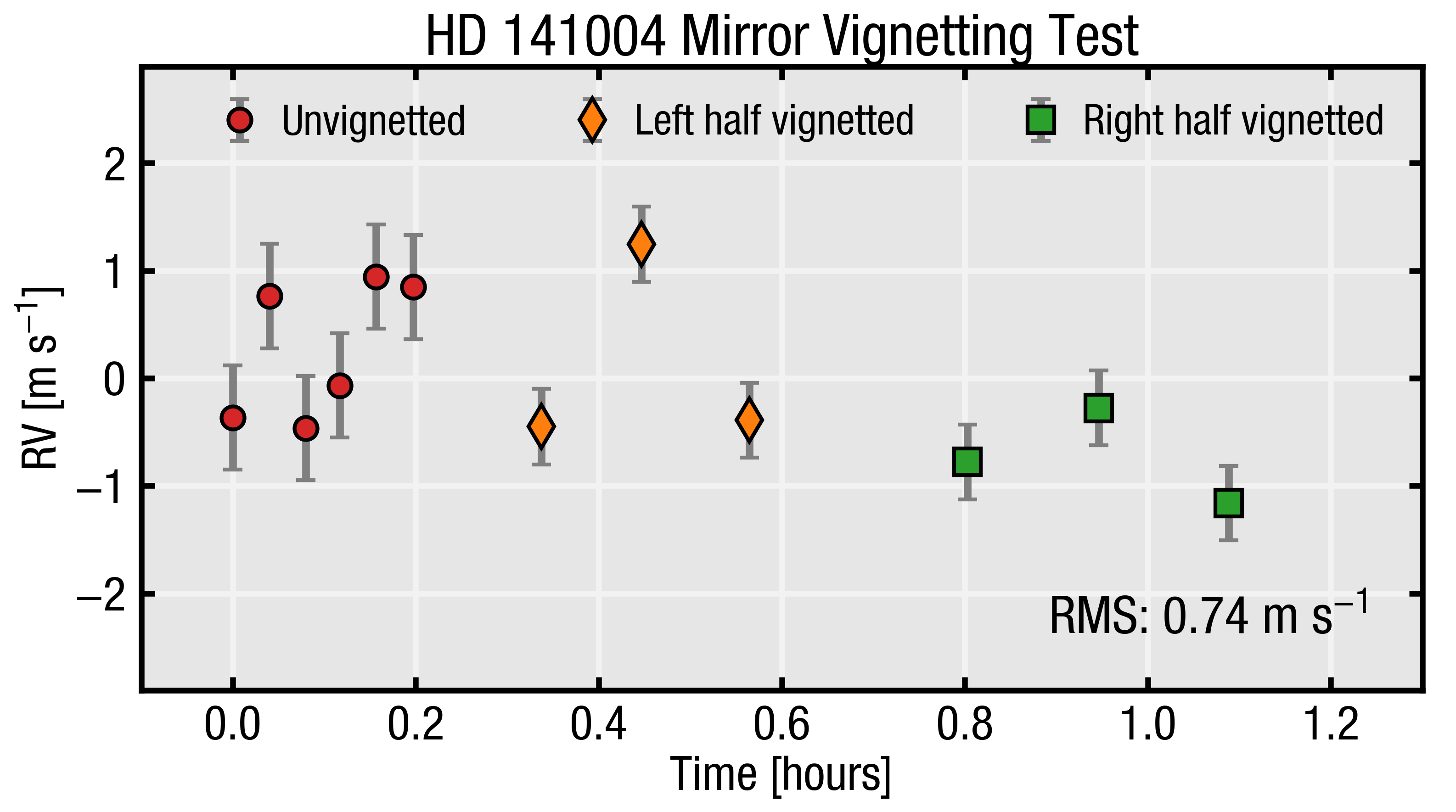}
\caption{Far field scrambling: On-sky testing of far-field scrambling performance. Error bars reflect photon-noise estimates only. HD141004 was observed with NEID under `standard' (full) mirror illumination (red circles), with one half of the WIYN primary mirror vignetted (orange diamonds), and the the other half vignetted (green squares). To maintain relatively constant signal-to-noise in the stellar spectra the exposure lengths were increased by $\sim$2x with half of the telescope pupil obscured. Within expected stellar jitter and photon-errors, we do not measure a statistically-significant offset between the three clusters of points.}
\label{fig:scrambling_far-field}
\end{figure}

To gauge the system sensitivity to input far-field variations, we performed a stress-test using stellar observations. To maximize the potential systematic signal induced by input pupil changes, we observed the bright star HD 141004 ($V\sim$~4.4) for several hours. During this time, we purposefully vignetted half of the WIYN primary pupil by partially closing the dome slit, and recorded a series of spectra with only half of the primary mirror fully illuminated, while we tracked the stellar flux using the exposure meter. We performed this test on a night with excellent seeing and transparency in order to ensure that the reduction in flux was deterministic and due to mirror obscuration rather than sky/cloud variations. After setting up the vignetting, the exposure times were triggered using the exposure meter to maintain comparable SNR throughout the test. The measured RVs for this test are shown in Figure~\ref{fig:scrambling_far-field}.

We found no significant difference between the derived target RVs for HD 141004 when the mirror was vignetted compared to when it was not. The RMS across the three cases shown in \autoref{fig:scrambling_far-field} is consistent with the expected stellar jitter from p-mode oscillations on the corresponding timescale \citep{gupta_detection_2022}. In general, pupil illumination variations arise from changes in seeing (which modulate the intensity pattern across the pupil), and injection angle variations (which move the pupil illumination). In either case, the expected variability is significantly less than our testing, and these observations represent an extremely harsh test of the NEID fiber scrambling capability which serves as a useful upper-limit for stability.

\section{Summary and Future Prospects}\label{sec:roadmap}

\begin{table}[ht]
\caption{The NEID instrumental error budget for terms relating to the optical fiber delivery system \citep{halverson_comprehensive_2016}. Terms that have been empirically explored in this study are marked with an asterisk. These were tested using internal calibration and/or on-sky experiments, which have provided upper limits on performance.\label{tab:err_budget}}
\begin{center}       
\begin{tabular}{lc}
\hline \hline
Fiber \& Illumination (total) & 9.8 \cms{} \\
\hline
Modal Noise (star) & 2.0 \cms{} \\
Modal Noise (calibration) & 2.0 \cms{} \\
Near-field scrambling* & 4.0 \cms{} \\
Far-field scrambling* & 4.0 \cms{} \\
Focal ratio degradation (star)* & 5.0 \cms{} \\
Focal ratio degradation (calibration)* & 5.0 \cms{} \\
Stray light + ghosts & 1.0 \cms{} \\
Fiber-fiber contamination & 1.0 \cms{} \\
Polarization variation & 2.0 \cms{} \\
\hline 
\end{tabular}
\tablenotetext{*}{Terms that have been empirically explored in this study.}
\end{center}
\end{table} 

\iffalse
\begin{figure}[!t] 
\center
\includegraphics[width=0.7\columnwidth]{err_budget_fib.png}
\caption{Condensed version of the NEID instrumental error budget for terms relating to the optical fiber delivery system \citep{halverson_comprehensive_2016}. Terms that have been empirically explored in this study are underlined. These have tested using via internal calibration testing and/or on-sky experiments, which have provided upper limits on performance. Continued testing is needed to reliably quantify these contributions, and to reduce each error contribution to the level necessary to support 10 {\cms} overall instrumental precision.}
\label{fig:err_budget}
\end{figure}
\fi

In this manuscript we have summarized the optical fiber-feed for NEID, its various sub-systems, and how the full-system Doppler error budget governed the requirements on the design and testing. While the fiber and illumination system is responsible for calibration instrumental noise, the error contribution from the system is not self-calibratable.  It is therefore extremely important to understand these errors and test them to reliably quantify their contribution to the overall instrument error budget, especially on the path to 10 \cms{} overall instrumental precision.  In \autoref{tab:err_budget}, we highlight a handful of error terms related to the fiber delivery system \citep[for multi-mode instruments;][]{halverson_comprehensive_2016}, along with their contribution to NEID's 27 \cms{} instrument error budget.

We have only begun to reach the instrumental capability necessary to isolate individual systematic error terms. As we move into the era of extreme precision RVs (EPRVs) which aim to approach 10 \cms{} or better, the performance of existing fiber systems must be interrogated even further. More detailed investigations that explore the subtle RV instability caused by far-field variations (such as pupil intensity changes, incident angle variations, focal shifts) are also critical for understanding the current systematic measurement floor. Moving forward, the magnitude of these issues should be quantified using precise calibration sources (i.e. non-stellar spectra) in controlled experiments, rather than relying on inferred upper-limits on performance based on on-sky data.  We note that there remain multiple obstacles towards improving the performance of next-generation fiber-feeds that support even tighter precision capabilities ($\sim$ \cms{}) in the quest for discovering an Earth analogue, and continued testing is needed to reliably quantify these contributions, and to reduce each error contribution to the level necessary to support 10 {\cms} overall instrumental precision. 

% In this study we have presented a blend of both approaches, but we also suggest further dedicated testing be done to bound specific error terms at the {\cms} level.

% \xx{} - add figure with subset of ORR NEID error budget, highlighting terms that need to be improved upon as we go towards 1 \cms{}. Talk about segmented ELTs, chief ray tolerance (Far field scrambling), along with pointing requirements for single mode fiber AO fed spectrographs? Perhaps a couple sentences on NIR spectrographs wouldn't be remiss (for handling stellar activity on younger stars, and of course M dwarf planets). Presumably the native 

% Open items on things which need to be improved upon
% Things that need better upper limits or constraints

% \begin{itemize}
%     \item proposed tests for modal noise (agitator on vs. off), LFC. Ideally using existing mechanical agitator in an on/off sequence
%     \item FRD experiments, following analysis in \cite{halverson_comprehensive_2016}, can be assessed using dedicated off-sky testing with calibration sources
%     \item adding a pupil stop to calibration source to mimick on-sky tests in a controlled manner with higher accuracy and precision
% \end{itemize}

\bibliography{references, References2}

\appendix
\section{Telescope Fiberhead measurements}

\begin{table*}[ht]
\caption{Final metrology for HR2 telescope bundle\label{tab:TelescopeHR}}
\begin{center}       
\begin{tabular}{ccccc}
\hline \hline
Fiber Name & x & y & Distance to Oct 1 & Distance to CFB 1 \\
           & mm & mm & mm & mm \\
\hline
\multicolumn{5}{l}{Fibers}  \\
Oct 1 & -0.1033 & -0.0781 & 0 & -\\
Oct 2 & 1.2095 & -0.8181 & 1.5070 & -\\
Oct 3 & -1.3940 & -0.8449 & 1.5012 & -\\
Oct 4 & -0.1117 & 1.4198 & 1.4980 & -\\
\multicolumn{5}{l}{Coherent Fiber Bundles}  \\
CFB 1 & 0.7748 & 0.4171 & 1.0082 & 0\\
CFB 2 & -0.0865 & -1.0897 & 1.0118 & 1.7357\\
CFB 3 & -0.9858 & 0.3899 & 0.9989 & 1.7608\\
\hline 
\end{tabular}
\tablenotetext{a}{The typical uncertainty for the fiber measurements is 0.1 $\mu$m, while that for the CFBs is 0.8 $\mu$m.}
\end{center}
\end{table*}

\begin{table*}[ht]
\caption{Final metrology for HE5 telescope bundle \label{tab:TelescopeHE}}
\begin{center}       
\begin{tabular}{ccccc}
\hline \hline
Fiber Name & x & y & Distance to Oct 1 & Distance to CFB 1 \\
           & mm & mm & mm & mm \\
\hline
\multicolumn{5}{l}{Fibers}  \\
Oct 1 & -0.0430 & -0.0266 & 0 & -\\
Oct 2 & 1.2643 & -0.7703 & 1.5041 & -\\
Oct 3 & -1.3336 & -0.7948 & 1.5019 & -\\
Oct 4 & -0.0582 & 1.4736 & 1.5003 & -\\
\multicolumn{5}{l}{Coherent Fiber Bundles}  \\
CFB 1 & 0.8175 & 0.4736 & 0.9953 & 0\\
CFB 2 & -0.0273 & -1.0334 & 1.0069 & 1.7277\\
CFB 3 & -0.9209 & 0.4303 & 0.9897 & 1.7389\\
\hline 
\end{tabular}
\tablenotetext{a}{The typical uncertainty for the fiber measurements is 0.1 $\mu$m, while that for the CFBs is 0.8 $\mu$m.}
\end{center}
\end{table*}

\begin{table}[ht]
\caption{Final throughput and FRD measurements for NEID fibers\label{tab:throughputfrd}}
\begin{center}   
\begin{tabular}{lcccc}
\hline \hline
Name                                  & Fiber                     & Type    & \multicolumn{1}{l}{Throughput \%} & \multicolumn{1}{l}{FRD (f/4 input)}  \\
                                      &                           &         & \multicolumn{1}{l}{}              & \multicolumn{1}{l}{Enclosed Energy \%} \\
                                      &                           &         & \multicolumn{1}{l}{}              & \multicolumn{1}{l}{at expected f/out} \\
\hline                                      
\textbf{Telescope HR 2$^a$}               & \multicolumn{1}{r}{Oct 1} & Science & 85.6                              & 96.4                                 \\
                                      & \multicolumn{1}{r}{Oct 2} &         & 85.6                              & 96.3                                 \\
                                      & \multicolumn{1}{r}{Oct 3} &         & 82.6                              & 94.3                                 \\
                                      & \multicolumn{1}{r}{Oct 4} & Sky     & 82.4                              & 96.4                                 \\
                                      \hline
\textbf{Telescope HE 5$^b$}               & \multicolumn{1}{r}{Oct 1} & Science & 87.8                              & 99.1                                 \\
                                      & \multicolumn{1}{r}{Oct 2} &         & 87.6                              & 98.5                                 \\
                                      & \multicolumn{1}{r}{Oct 3} &         & 86.8                              & 97.4                                 \\
                                      & \multicolumn{1}{r}{Oct 4} & Sky     & 87.6                              & 98.2                                 \\
                                      \hline
\textbf{Bifurcated Fiber 2$^c$}           & Fiber C                   & Solar   & 93.4                              & 70                                   \\
                                      & Fiber D                   & Cal     & 94.4                              & 89                                   \\
                                      \hline
\textbf{Cal Fiber Patch$^d$}              & PSR9                      &         & 92.8                              & 97                                   \\
\hline
\textbf{Instrument Fiber}             &                           &         & \multicolumn{1}{l}{}              & \multicolumn{1}{l}{}                 \\
\hline
\textbf{Testing the vacuum feedthrough$^e$} & HR1                       & Cal     & 90.4                              & 97.9                                 \\
                   & HR2                       & Science & 90.9                              & -                 \\
                                      & HR3                       & Sky     & 91.5                              & 98.2                                 \\
                                      & HE1                       & Science & 90.9                              & 98.5                                 \\
                                      & HE2                       & Sky     & 91.5                              & 98.3                                 \\
                                      \hline
\textbf{After final polish of the puck$^f$}           & HR1                       & Cal     & 96                                & 86.3                                 \\
                                      & HR2                       & Science & 95.8                              & 91.8                                 \\
                                      & HR3                       & Sky     & 96                                & 89.1                                 \\
                                      & HE1                       & Science & 94.6                              & 89.3                                 \\
                                      & HE2                       & Sky     & 94.5                              & 90.8                                 \\
                                      \hline
\textbf{After L1 singlet$^g$}               & HR1                       & Cal     & 86.1                              & 94.1                                 \\
                                      & HR2                       & Science & 92.4                              & 95.9                                 \\
                                      & HR3                       & Sky     & 88.1                              & 95.2                                 \\
                                      & HE1                       & Science & 95                                & 96.1                                 \\
                                      & HE2                       & Sky     & 92.9                              & 95.4                                 \\
                                      \hline
\textbf{After L1 singlet + optics tube$^g$}              & HR1                       & Cal     & 93.3                              & 87.3                                 \\
                                      & HR2                       & Science & 97.9                              & 89.6                                 \\
                                      & HR3                       & Sky     & 86.7                              & 87.2                                 \\
                                      & HE1                       & Science & 95.6                              & 87.8                                 \\
                                      & HE2                       & Sky     & 91.3                              & 88.6       \\                         \hline
\end{tabular}
\end{center}
\tablenotetext{a}{Final HR telescope fiberhead}
\tablenotetext{b}{Final HE telescope fiberhead}
\tablenotetext{c}{Used to input light from the calibration bench \citep[or the solar telescope;][]{lin_observing_2022} to the port adapter into the HR fiber}
\tablenotetext{d}{Calibration bench to the ball lens, for simultaneous calibration of stellar observations}
\tablenotetext{e}{We test the fiber stresses in the vacuum feedthrough by potting the fibers at both ends in separate brass ferrules, and testing their performance.}
\tablenotetext{f}{Compared to the previous step, this includes a 2 m octagonal fiber + spliced to the circular instrument fiber, which is terminated at the fused silica puck}
\tablenotetext{g}{Compared to the previous step, this includes a fused silica singlet lens glued to the fused silica puck}
\tablenotetext{h}{Compared to the previous step, this includes series of lenses which serve as a f/$\#$ converter \citep[optics tube][]{schwab_neid_2020}.}
\end{table}

\section{Acknowledgements}

NEID is funded by NASA through JPL by contract 1547612.

The Center for Exoplanets and Habitable Worlds is supported by Penn State and the Eberly College of Science.

Based in part on observations at Kitt Peak National Observatory, NSF’s NOIRLab, managed by the Association of Universities for Research in Astronomy (AURA) under a cooperative agreement with the National Science Foundation. The authors are honored to be permitted to conduct astronomical research on Iolkam Du’ag (Kitt Peak), a mountain with particular significance to the Tohono O’odham.
 
Data presented herein were obtained at the WIYN Observatory from telescope time allocated to NN-EXPLORE through the scientific partnership of the National Aeronautics and Space Administration, the National Science Foundation, and the National Optical Astronomy Observatory.

Deepest gratitude to Zade Arnold, Joe Davis, Michelle Edwards, John Ehret, Tina Juan, Brian Pisarek, Aaron Rowe, Fred Wortman, the Eastern Area Incident Management Team, and all of the firefighters and air support crew who fought the recent Contreras fire. Against great odds, you saved Kitt Peak National Observatory.

We thank the NEID Queue Observers and WIYN Observing Associates for their skillful execution of our NEID observations.

This research has made use of 
%the SIMBAD database, operated at CDS, Strasbourg, France, 
%and 
NASA's Astrophysics Data System Bibliographic Services.

GS acknowledges support provided by NASA through the NASA Hubble Fellowship grant HST-HF2-51519.001-A awarded by the Space Telescope Science Institute, which is operated by the Association of Universities for Research in Astronomy, Inc., for NASA, under contract NAS5-26555.

\end{document}